\documentclass[11pt,epsf]{article}
\DeclareMathAlphabet{\scr}{U}{rsfs}{m}{n}
\usepackage{latexsym,color}
\usepackage{epsfig}
\usepackage[mathscr]{eucal}
\usepackage{amsfonts}
\usepackage{amscd}
\usepackage{cite}
\usepackage{array}
\usepackage{amssymb}
\usepackage{colordvi}
\usepackage[centertags]{amsmath}
\usepackage{enumerate}
\usepackage{graphicx}
\usepackage{booktabs}
\usepackage{theorem}
\usepackage[footnotesize]{caption}
\usepackage{soul}

\newcommand{\newc}{\newcommand}
\newc{\be}{\begin{equation}}
\newc{\ee}{\end{equation}}
\newc{\bea}{\begin{eqnarray}}
\newc{\eea}{\end{eqnarray}}
\newc{\ol}{\overline}
\newc{\wt}{\widetilde}
\newc{\bs}{\boldsymbol}
\newc{\m}{\mathcal}
\newc{\la}{\langle}
\newc{\ra}{\rangle}

\newcommand{\non}{\nonumber}
\newcommand{\beq}{\begin{eqnarray}}
\newcommand{\eeq}{\end{eqnarray}}
\newcommand{\s}{\smallskip}

\newcommand{\msta}{m_{\tilde{t}_1}}
\newcommand{\mstb}{m_{\tilde{t}_2}}

\newcommand{\mak}{M_{A_k}}
\newcommand{\DRb}{\overline{\text{DR}}}

\newcommand{\MSb}{\overline{\text{MS}}}

\newcommand{\bc}{\begin{center}}
\newcommand{\ec}{\end{center}}

\newcommand{\gev}{~\text{GeV}}
\newcommand{\mev}{~\text{MeV}}

\newcommand{\ba}{\begin{array}}
\newcommand{\ea}{\end{array}}

\begin{document}

\title{
\vspace*{-3cm}
\phantom{h} \hfill\mbox{\small KA-TP-11-2015}
\\[1cm]
\textbf{Next-to-Leading Order NMSSM Decays with CP-odd Higgs Bosons and Stops}}

\date{}
\author{
J.~Baglio$^{1\,}$\footnote{E-mail: \texttt{julien.baglio@kit.edu}},
C.O.~Krauss$^{2\,}$\footnote{E-mail: \texttt{carla.olivia.k@googlemail.com}}, 
M. M\"{u}hlleitner$^{2\,}$\footnote{E-mail: \texttt{margarete.muehlleitner@kit.edu}},
K. Walz$^{2\,}$\footnote{E-mail: \texttt{kathrin.walz@kit.edu}}
\\[9mm]
{\small\it
$^1$Institute for Theoretical Physics, T\"ubingen University,} \\
{\small \it Auf der Morgenstelle 14, 72076 T\"ubingen, Germany}\\[3mm]
{\small\it
$^2$Institute for Theoretical Physics, Karlsruhe Institute of Technology,} \\
{\small\it Wolfgang-Gaede-Str. 1, 76131 Karlsruhe, Germany}
}

\maketitle

\begin{abstract}
\noindent
We compute the full next-to-leading order supersymmetric (SUSY)
electroweak (EW) and SUSY-QCD corrections to the decays of CP-odd
NMSSM Higgs bosons into stop pairs. In our numerical analysis we also present the
decay of the heavier stop into the lighter stop and an NMSSM CP-odd
Higgs boson. Both the EW and the SUSY-QCD corrections are found to be
significant and have to be taken into account for a proper prediction
of the decay widths. 
\end{abstract}
\thispagestyle{empty}
\vfill
\newpage
\setcounter{page}{1}

\section{Introduction}
The announcement of the discovery of a new boson by the LHC
experiments ATLAS \cite{:2012gk} and CMS \cite{:2012gu} has marked a
milestone for particle physics. While the properties of this particle 
are consistent with the SM predictions, the uncertainties in the
experimental data still leave enough room for interpretations in
extensions beyond the SM. Among these, supersymmetric (SUSY) models
\cite{susy} certainly belong to the best motivated and most intensely
studied ones. In particular the Next-to-Minimal Supersymmetric Extension
(NMSSM)~\cite{nmssm} provides with the introduction of an additional
complex superfield $\hat{S}$ a dynamical solution to
the $\mu$ problem~\cite{muproblem} when the singlet field acquires a
non-vanishing vacuum expectation value.  Because of new contributions
to the quartic coupling $\lambda$, with which $\hat{S}$ couples to the
Higgs doublet superfields $\hat{H}_u$ and $\hat{H}_d$, the tree-level
mass value of the lighter MSSM-like Higgs boson is enhanced. In
consequence less important radiative corrections are required to shift
the mass value to the measured value of 125~GeV and therefore smaller
stop masses and/or mixing are necessary, so that the fine-tuning is reduced
\cite{finetune,ournmssmpapers}. \s

After electroweak symmetry breaking
(EWSB) the Higgs sector of the NMSSM consists of seven physical Higgs
bosons. In the CP-conserving case, which we assume to be valid here,
these are three neutral CP-even, two neutral CP-odd and two charged
Higgs bosons. The discovery of all Higgs particles is challenging
though not impossible at the high-energy option of the LHC. In
\cite{King:2014xwa} we investigated the discovery prospects for the
NMSSM Higgs bosons during the 13 TeV run of the LHC and gave benchmark
scenarios that feature Higgs-to-Higgs decays.\footnote{For other recent
  studies on NMSSM Higgs boson phenomenology, see
  Refs.~\cite{ournmssmpapers,brchanges,Gunion:2012gc,Gunion:2012he,Kang:2013rj,Cerdeno:2013cz,Badziak:2013bda,Munir:2013wka,Nhung:2013lpa,Ellwanger:2013ova,Cerdeno:2013qta,Beskidt:2013gia,Choi:2013lda,Kozaczuk:2013spa,
    Cao:2013gba,Jia-Wei:2013eea,Huang:2013ima,Munir:2013dya,Belanger:2014roa,Beskidt:2014oea,Ellwanger:2014dfa,Ellwanger:2014hia,Das:2014fha,Cao:2014kya,Bomark:2014gya,Ellwanger:2014hca,Moretti:2015bua,Buttazzo:2015bka,Potter:2015wsa}.}
If kinematically allowed also decays into supersymmetric particles can
become important, as is well known for the Minimal
Supersymmetric Extension of the SM (MSSM)
\cite{Djouadi:1996pj,Djouadi:1996mj}. The one-loop SUSY-QCD corrections to the 
decays into stops and sbottoms of the MSSM Higgs bosons have been calculated
in \cite{Bartl:1997yd,Arhrib:1997nf,Eberl:1999he} and can change the decay widths
by more than 50\%, especially near the threshold. The SUSY-QCD
corrections have been reanalyzed in
\cite{Accomando:2011jy}. The full electroweak  
(EW) one-loop corrections to the pseudoscalar decays into squarks have
been provided in \cite{Weber:2003eg} and have turned out to be
significant. Equally, the decays of heavy squarks into lighter ones
and a Higgs boson can dominate in a wide range of the MSSM parameter
space due to the large Yukawa couplings and stop and sbottom mixing
\cite{Bartl:1998xk}. The SUSY-QCD corrections at next-to-leading order
(NLO) are of the order of a few ten percent \cite{Bartl:1998xp} and mostly
negative. The one-loop EW corrections to the decays with a
pseudoscalar in the final state are significant
\cite{Weber:2003eg}. The full one-loop corrections for the complex MSSM
have been discussed in \cite{Heinemeyer:2010mm}. \s

The proper interpretation of the experimental data and, once SUSY has
been discovered, the aim to pin down the underlying model and distinguish {\it
  e.g.}~the NMSSM from the MSSM, require precise predictions both for
the parameters of the model and for the observables like {\it e.g.}~NMSSM Higgs
boson production and decay rates \cite{Kramer:2015pea}. The higher
order corrections to the 
CP-conserving NMSSM Higgs boson masses and self-couplings have been given in
\cite{Ellwanger:1993hn,Elliott:1993ex,Elliott:1993uc,Elliott:1993bs,Pandita:1993tg,Ellwanger:2005fh,Degrassi:2009yq,Staub:2010ty,Ender:2011qh,Goodsell:2014pla,Goodsell:2015ira}
and \cite{Nhung:2013lpa}, respectively. 
Additionally, there are several codes available for the evaluation of the NMSSM mass
spectrum from a user-defined input at a user-defined scale, like {\tt NMSSMTools}
\cite{Ellwanger:2004xm,Ellwanger:2005dv,Ellwanger:2006rn} which can be
interfaced with {\tt SOFTSUSY} \cite{Allanach:2001kg,Allanach:2013kza}, the 
interface of {\tt SARAH}
\cite{Staub:2010jh,Staub:2012pb,Staub:2013tta,Goodsell:2014bna,Goodsell:2014pla}
with {\tt SPheno} \cite{Porod:2003um,Porod:2011nf}, and finally {\tt
  SARAH} which has been interfaced with the recently published package
{\tt FlexibleSUSY} \cite{Athron:2014yba,Athron:2014wta}. Recently,
{\tt NMSSMTools} has been extended to include also the CP-violating NMSSM
\cite{Domingo:2015qaa}. 
In our Fortran package {\tt NMSSMCALC} \cite{Baglio:2013iia} we have
included in the CP-conserving and CP-violating NMSSM the full
one-loop and the order ${\cal O}(\alpha_t \alpha_s)$ corrections to the NMSSM
Higgs boson masses \cite{Ender:2011qh,Graf:2012hh,Muhlleitner:2014vsa}
and the state-of-the-art higher order corrections to the decays. These
include in the CP-conserving case Higgs decays into stops and sbottoms
the SUSY-QCD 
corrections of \cite{Accomando:2011jy} which have been adapted from
the MSSM to the NMSSM case. Very recently, neutral Higgs production
through gluon fusion and bottom-quark annihilation including higher
order corrections has been discussed in \cite{Liebler:2015bka}. \s

With this work we take another step in improving the predictions for
the NMSSM Higgs sector. We provide both the NLO SUSY-QCD and the full one-loop
EW corrections to the decays of a pseudoscalar NMSSM Higgs
boson into stops as well as to the decays of the heavier stop into the 
lighter one and a pseudoscalar. \s

The paper is organized as follows. In section \ref{sec:nmssm} we
introduce the NMSSM Higgs sector and set our notation. The tree-level
decay of a pseudoscalar Higgs into stops is discussed in section
\ref{sec:treedec}, before we present in section \ref{sec:susyqcd} the order
${\cal O}(\alpha_s)$ SUSY-QCD corrections and in section \ref{sec:ewcorr}
the one-loop EW corrections. The numerical analysis, including the
discussion of heavier stop decays into a pseudoscalar and lighter stop
final state, is performed in section 
\ref{sec:numerical}. We conclude in section \ref{sec:concl}.

\section{The NMSSM Higgs Sector \label{sec:nmssm}}
The NMSSM Higgs potential is obtained from the NMSSM superpotential,
which we assume to be scale invariant, the soft SUSY breaking terms
and the $D$-term contributions, which are the same as in the MSSM. In
terms of the superfields $\hat{H}_u$ and $\hat{H}_d$, coupling to the
up- and down-type quarks, respectively, and the singlet superfield
$\hat{S}$ the superpotential reads
\beq
W_{NMSSM} = W_{MSSM} - \epsilon_{ij} \lambda \hat{S} \hat{H}^i_d
\hat{H}^j_u + \frac{1}{3} \kappa \hat{S}^3 \;,
\eeq
where $i,j=1,2$ are the $SU(2)_L$ indices and we have introduced
the totally antisymmetric tensor $\epsilon_{ij}$ with
$\epsilon_{12}=1$. Working in the 
CP-invariant NMSSM, the dimensionless parameters $\lambda$ and
$\kappa$ are chosen to be real. The MSSM superpotential $W_{MSSM}$ is
given by 
\beq
W_{MSSM} = \epsilon_{ij} [y_e \hat{H}^i_d \hat{L}^j \hat{E}^c + y_d
\hat{H}_d^i \hat{Q}^j \hat{D}^c - y_u \hat{H}_u^i \hat{Q}^j \hat{U}^c] 
\label{eq:mssmsuperpot} \;,
\eeq
with the quark and lepton superfields and their charge conjugates,
indicated by the superscript $c$, denoted by $\hat{Q}, \hat{U}^c,
\hat{D}^c, \hat{L}$ an $\hat{E}^c$. The color and generation indices
have been suppressed in Eq.~(\ref{eq:mssmsuperpot}). We neglect
generation mixing of the quarks, so that the phases of the Yukawa
couplings $y_d$, $y_u$ and $y_e$, which in general are complex, can be
reabsorbed by a redefinition of the quark fields. The soft SUSY
breaking NMSSM Lagrangian involving the Higgs doublet and singlet
component fields $H_u$, $H_d$ and $S$ reads
\beq
\mathcal L_{soft} = {\cal L}_{soft,\, MSSM} - m_S^2 |S|^2 +
(\epsilon_{ij} \lambda 
A_\lambda S H_d^i H_u^j - \frac{1}{3} \kappa
A_\kappa S^3 + h.c.) \;, \label{eq:nmssmsoft}
\eeq
with the soft SUSY breaking MSSM Lagrangian
\beq
{\cal L}_{soft, \, MSSM} &=& -m_{H_d}^2 H_d^\dagger H_d - m_{H_u}^2
H_u^\dagger H_u -
m_{{\tilde{Q}}}^2 \tilde{Q}^\dagger \tilde{Q} - m_{{\tilde{L}}}^2
\tilde{L}^\dagger \tilde{L} - m_{{\tilde{U}}_R}^2 \tilde{u}_R^* 
\tilde{u}_R - m_{{\tilde{D}}_R}^2 \tilde{d}_R^* \tilde{d}_R 
\nonumber \\\nonumber
&& - m_{{\tilde{E}}_R}^2 \tilde{e}_R^* \tilde{e}_R - (\epsilon_{ij} [y_e A_E H_d^i
\tilde{L}^j \tilde{e}_R^* + y_d
A_D H_d^i \tilde{Q}^j \tilde{d}_R^* - y_u A_U H_u^i \tilde{Q}^j
\tilde{u}_R^*] + h.c.) \\
&& -\frac{1}{2}(M_1 \tilde{B}\tilde{B} + M_2
\tilde{W}_k\tilde{W}_k + M_3 \tilde{G}\tilde{G} + h.c.) \;.
\label{eq:mssmsoft}
\eeq
The gaugino fields are denoted by $\tilde{B}$, $\tilde{W}_k$
($k=1,2,3$) and $\tilde{G}$, and the left-handed squarks and sleptons
are arranged in doublets denoted by
$\tilde{Q}=(\tilde{u}_L,\tilde{d}_L)^T$, $\tilde{L}=(\tilde{\nu}_L,
\tilde{e}_L)^T$ while the right-handed fields are denoted by $\tilde
u_R$, $\tilde d_R$ and $\tilde e_R$. The soft SUSY breaking
mass parameters $m_{X_{(R)}}^2$ of the scalar fields
$X=S,H_d,H_u,\tilde{Q},\tilde{U},\tilde{D},\tilde{L},\tilde{E}$  
are real, while the gaugino mass parameters $M_1,M_2$ and $M_3$ and
the soft SUSY breaking trilinear couplings $A_Y$
($Y=\lambda,\kappa,U,D,E$)  are in 
general complex. In the CP-conserving case assumed here, they are, 
however, real. Again, the respective quark and lepton superfields are
understood to refer to all three fermion generations. 
Note that we have set soft SUSY breaking
terms linear and quadratic in the singlet field $S$ to zero. \s

After expanding the Higgs fields about their vacuum expectation values
(VEVs) $v_u$, $v_d$ and $v_s$, chosen to be real and positive,
\beq
H_d = \left( \begin{array}{c} (v_d + h_d + i a_d)/\sqrt{2} \\
   h_d^- \end{array} \right) \;, \quad
H_u = \left( \begin{array}{c} h_u^+ \\ (v_u + h_u + i a_u)/\sqrt{2}
 \end{array} \right) \;, \quad
S= \frac{v_s+h_s+ia_s}{\sqrt{2}} \;,
\eeq
the Higgs mass matrices for the three scalar, two pseudoscalar and  the
charged Higgs bosons can be derived from the tree-level scalar
potential. The mass matrix decomposes into two mass matrices for the
CP-even and the CP-odd Higgs fields. The squared $3 \times 3$ mass
matrix $M_S^2$ for the CP-even Higgs fields can be diagonalized
through a rotation matrix ${\cal R}^S$ which yields the CP-even mass
eigenstates $H_i$ ($i=1,2,3$) as
\beq
(H_1, H_2, H_3)^T = {\cal R}^S (h_d,h_u,h_s)^T \;.
\label{eq:scalarrot}
\eeq
The $H_i$ are ordered by ascending mass, $M_{H_1} \le M_{H_2} \le
M_{H_3}$. In order to obtain the CP-odd mass eigenstates $A_1$, $A_2$ and
the massless Goldstone boson $G$ first a rotation
${\cal R}^G$ to separate $G$ is applied, and then a 
rotation ${\cal R}^P$ to obtain the mass eigenstates
\beq
(A_1,A_2,G)^T = {\cal R}^P (a,a_s,G)^T = {\cal R}^P {\cal R}^G (a_d,a_u,a_s)^T \;,
\label{eq:pseudorot}
\eeq
which are ordered such that $M_{A_1} \le M_{A_2}$.  \s

The minimisation of the Higgs potential $V$ requires the terms linear
in the Higgs fields to vanish in the vacuum. The corresponding
coefficients, which are called tadpoles, therefore have to be
zero. The tadpole conditions for the CP-even fields can be exploited
to replace $m_{H_u}^2,m_{H_d}^2$ and $m_S^2$ by the tadpole
parameters $t_{h_d}, t_{h_u}$ and $t_{h_s}$. Replacing the $SU(2)_L$
and $U(1)_Y$ gauge couplings $g$ 
and $g'$ and the VEVs $v_u$ and $v_d$ by the electric charge $e$, the
gauge boson masses $M_W$, $M_Z$ and by $\tan\beta$, the tree-level
NMSSM Higgs sector can then be parameterized by the twelve parameters
\beq
t_{h_u}, \ t_{h_d}, \ t_{h_s}, \ e , \ M_W^2, \ M_Z^2, \ 
\lambda, \ \kappa, \ A_{\lambda}, \ A_{\kappa}, \
\tan \beta =\langle H_u^0 \rangle / \langle H_d^0 \rangle \quad \mathrm{and}
\quad \mu_\mathrm{eff} = \lambda \langle S \rangle\; .
\eeq 
The VEVs of the neutral components of the Higgs fields are denoted by
the brackets around the corresponding fields. The sign
conventions for $\lambda$ and $\tan\beta$ are chosen such that they
are positive. The $\kappa$, $A_\lambda$, $A_\kappa$ and
$\mu_{\mathrm{eff}}$ on the other hand can have both signs. Note also,
that the parameter $A_\lambda$ can be traded for the charged Higgs
boson mass $M_{H^\pm}$, which we will do in the
following. From now 
on we will drop the subscript '$\text{eff}$'. Note that
the inclusion of higher order corrections requires also the soft SUSY
breaking mass terms for the scalars and the gauginos as well as the
trilinear soft SUSY breaking couplings. 

\section{The Tree-level Decay Width \label{sec:treedec}}
We start by discussing the tree-level decay width of a pseudoscalar
Higgs boson $A_i$ ($i=1,2$) into a pair of stops. 
The stop mass matrix in the interaction basis $(\tilde{t}_L,\tilde{t}_R)$ reads
\beq
{\cal M}^2_{\tilde{t}} = \left( \begin{array}{cc} m_{LL}^2 & m_{LR}^2
    \\ m_{RL}^2 & m_{RR}^2 \end{array} \right) \;,
\eeq
with
\beq
m_{LL}^2 &=& m_{\tilde{Q}}^2 + m_t^2 + M_Z^2 \cos 2\beta \left( \frac{1}{2} -
\frac{2}{3} \sin^2 \theta_W \right) \\
m_{RR}^2 &=& m_{\tilde{t}_R}^2 + m_t^2 + \frac{2}{3 }M_Z^2 \cos 2\beta \sin^2
\theta_W \\
m_{LR}^2 = m_{RL}^2 &=& m_t (A_t - \mu \cot\beta) \;,
\eeq
in terms of the soft SUSY breaking mass parameters $m_{\tilde{Q}}$ and
$m_{\tilde{t}_R}$, the soft SUSY breaking trilinear coupling $A_t$, the
higgsino mixing parameter $\mu$, the top and the $Z$ boson
masses $m_t$ and $M_Z$, the mixing angle $\beta$ and the Weinberg
angle $\theta_W$. The $\mu$ parameter is generated dynamically in the
NMSSM and given by 
\beq
\mu = \frac{\lambda v_s}{\sqrt{2}} \;.
\eeq
The stop mass matrix is diagonalized by
\beq
{\cal R}^{\tilde{t}} = \left( \begin{array}{cc} \cos
    \theta_{\tilde{t}} & \sin \theta_{\tilde{t}} \\ -\sin
    \theta_{\tilde{t}} & \cos \theta_{\tilde{t}}  \end{array} \right)
\eeq
yielding the stop mass eigenstates $\tilde{t}_i$ ($i=1,2$) as
\beq
\tilde{t}_i =  {\cal R}^{\tilde{t}}_{is} \tilde{t}_s \;,
\eeq
where $s=L,R$ and for the squark masses we have $m_{\tilde{t}_1} <
m_{\tilde{t}_2}$ by convention. The mixing angle $\theta_{\tilde{t}}$
and the squark masses are given by
\beq
\tan \theta_{\tilde{t}} = \frac{2
  m_{LR}^2}{m_{LL}^2-m_{RR}^2-\sqrt{(m_{LL}^2-m_{RR}^2)^2 + 4
    m_{LR}^4}}
\eeq
and
\beq
m_{\tilde{t}_{1,2}}^2 = \frac{1}{2} \left[ m_{LL}^2 + m_{RR}^2
\mp \sqrt{(m_{LL}^2-m_{RR}^2)^2+4 m_{LR}^4} \right] \;.
\eeq
In case of the pseudoscalar only the coupling to two different stop
mass eigenstates is non-vanishing. For the tree-level decay width
$\Gamma^{\text{LO}}$ we have
\beq
\Gamma^{\text{LO}} (A_i \to \tilde{t}_1 \tilde{t}_2) = \frac{3
  \lambda^{1/2} (M_{A_i}^2, m_{\tilde{t}_1}^2,
m_{\tilde{t}_2}^2 )}{8 \pi M_{A_i}^3} 
\left|\sum_{j=1}^2 {\bf Z}_{ij} G_{A_j}^{12} \right|^2
\;,
\label{eq:treegam}
\eeq
where $\lambda (x,y,z) = (x-y-z)^2-4yz$ is the two-body phase space
function and the coupling $G_{A_j}^{12}$ ($j=1,2$) of the pseudoscalar
$A_j$ to the stops reads
\beq
G_{A_j}^{12} =  \frac{g m_t}{2 M_W} \left[ \left( \frac{A_t}{\tan\beta} + \mu \right)
  {\cal R}^P_{j1} + \frac{\lambda v}{\sqrt{2}} \frac{1}{\tan\beta}
  {\cal R}^P_{j2} \right]
\eeq
where $v$ is the VEV given by $v^2=v_u^2 +v_d^2$ and ${\cal R}^P_{jk}$
are the elements of the rotation matrix defined 
in Eq.~(\ref{eq:pseudorot}). In particular, they are the tree-level
mixing matrix elements. In the kinematics of the decay, however, {\it
  i.e.}~for the external Higgs field, we use the two-loop corrected
Higgs boson masses at order ${\cal O}(\alpha_t \alpha_s)$, which
include the full EW corrections at one-loop order. The
renormalization of the Higgs fields and the computation of the mass
corrections have been described in
Refs.~\cite{Ender:2011qh,Graf:2012hh,Muhlleitner:2014vsa}. 
We follow the conventions of these papers, to which we refer the
reader for more details. In order to ensure the on-shell properties
of the external Higgs field, which in the calculation of the Higgs
mass corrections has been renormalized in the
mixed on-shell--$\overline{\mbox{DR}}$ scheme, the finite wave function
renormalization factors ${\bf Z}$ \cite{Frank:2006yh} have to be taken
into account. The application of the factor ${\bf Z}$ to the
tree-level matrix ${\cal R}^P$ (in $G_{A_j}^{12}$) leads to the
rotation matrix ${\cal R}^{P,l}$ which, modulo the Goldstone boson
$G$, rotates the interaction eigenstates $a$ and $a_s$ to the loop
corrected mass eigenstates $A_1$ and $A_2$, {\it
  cf.}~\cite{Ender:2011qh},
\beq
{\cal R}^{P,l}_{il} = {\bf Z}_{ij} {\cal R}^P_{jl} \;, \qquad i,j=
A_1, A_2,G \;, \quad l=a,a_s,G \;.
\eeq
These and the loop-corrected masses are taken 
from the Fortran code {\tt NMSSMCALC} \cite{Baglio:2013iia}, in which
we choose the on-shell (OS) renormalization for the top/stop sector in order to
be in accordance with the renormalization scheme chosen later on both
in the SUSY-QCD corrections and in the electroweak corrections. \s

In Eq.~(\ref{eq:treegam}) we
have summed over both possible final states $\tilde{t}_1
\tilde{t}_2^*$ and $\tilde{t}_1^* \tilde{t}_2$. In the MSSM, {\it
  i.e.}~leaving out the singlet contribution $\propto \lambda$ in the
coupling $G^{12}_{A_i}$, the decay width is
proportional to $m_t^2 (\mu + A_t / \tan\beta)^2 /
M_{A_i}$. For small values of $\tan\beta$ and not too heavy pseudoscalars, the
decay into stops can compete with and even dominate over the decays
into top quarks and into charginos and neutralinos. In the NMSSM, this
statement has to be taken with caution, however, as the singlet component in
the coupling $G^{12}_{A_i}$, depending on the scenario, can come with
both signs and hence increase or decrease the decay width. 

\section{SUSY-QCD Corrections \label{sec:susyqcd}}
The SUSY-QCD corrections for the NMSSM pseudoscalar $A_i$ decay width differ
from the ones of the MSSM 
\cite{Bartl:1997yd,Arhrib:1997nf,Eberl:1999he} solely in the
tree-level coupling to the stops $G_{A_i}^{12}$. We shortly repeat
them here for completeness and in order to introduce our
renormalization scheme. \s 
 
\begin{figure}[t]
\begin{center}
\includegraphics[width=0.8\textwidth]{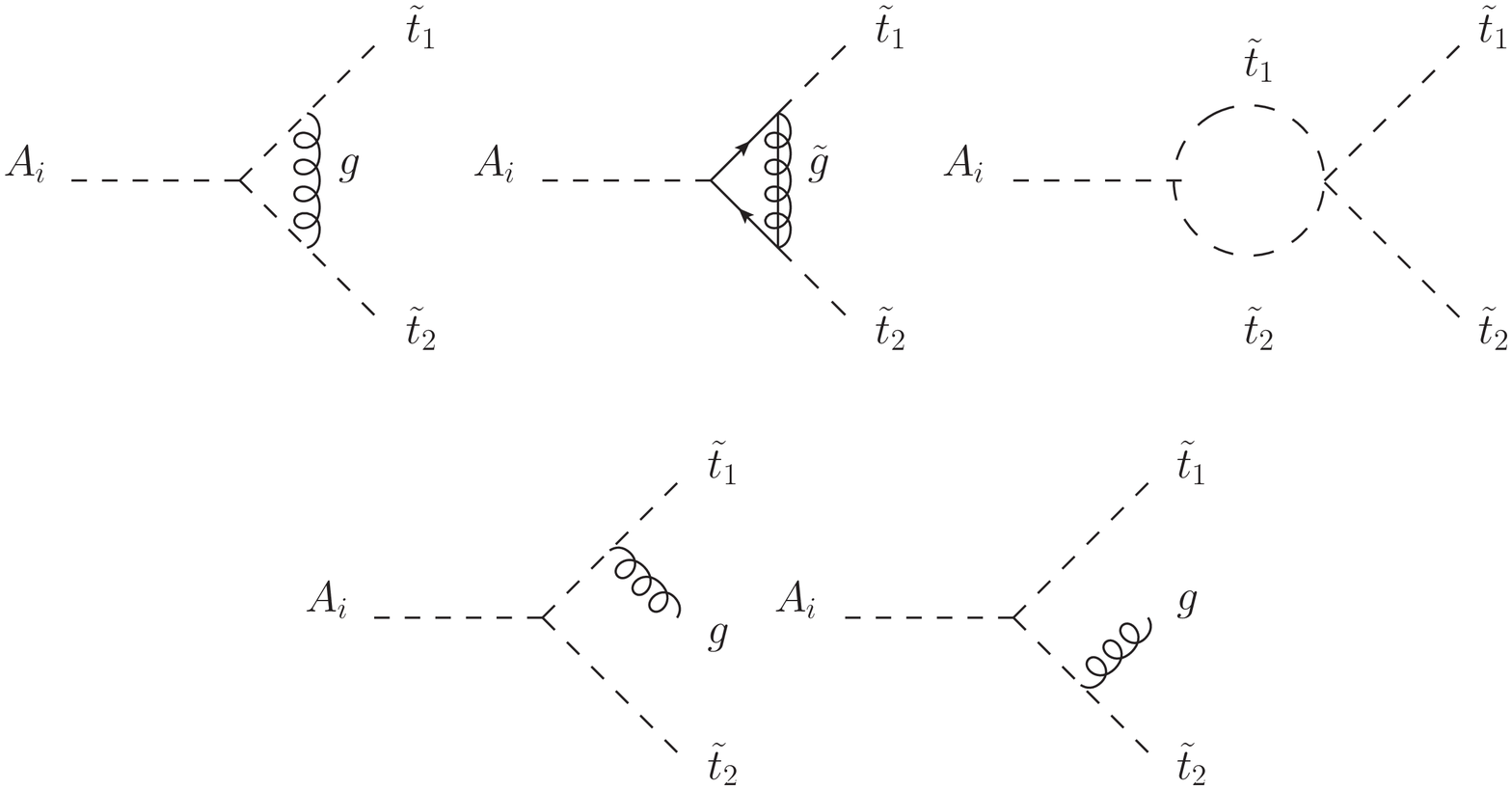}
\end{center}
\caption{Diagrams contributing at NLO SUSY-QCD to the virtual (upper
  row) and real corrections (lower row) of the decay $\Gamma (A_i \to
  \tilde{t}_1 \tilde{t}_2)$.} \label{fig:nlosusyqcd} 
\end{figure}

The virtual corrections at order ${\cal O} (\alpha_s)$ to the
pseudoscalar Higgs decays into stops consist of  loop diagrams
with a gluon, respectively, gluino exchanged in the
$A_i \tilde{t}_1 \tilde{t}_2$ vertex and of a contribution involving the
four-squark vertex, {\it cf.}~Fig.~\ref{fig:nlosusyqcd} (upper). Note
that the mixing contributions due to off-diagonal self-energies are
absent in the case of pseudoscalar Higgs decays, as $A_i$ only couples
to different stop mass eigenstates. The computation of the virtual
diagrams leads to ultraviolet (UV) and infrared (IR) divergences. We
work in dimensional reduction \cite{dimred}, which preserves SUSY at the
one-loop level. The fields and couplings are then treated in 4
dimensions while the loop integral is performed in $D=4-2\epsilon$
dimensions. The UV divergences appear as poles in $\epsilon$ and are
canceled by the wave function counterterms and the counterterm
renormalizing the $A_i \tilde{t}_1\tilde{t}_2$ interaction. The infrared
divergences are regularized by the introduction of a fictitious gluon mass
$\zeta$. The IR divergences left over after renormalization are
canceled after adding the real corrections. These consist of the
radiation of an additional gluon off the final state stops and are
shown in Fig.~\ref{fig:nlosusyqcd} (lower). \s

The one-loop corrected decay amplitude can be written as
\beq
\Gamma^{\text{NLO}}_{\text{QCD}} = \Gamma^{\text{LO}} +
\Gamma^{(1)}_{\text{QCD}} \;, \label{eq:nloqcdgam}
\eeq
with
\beq
\Gamma^{(1)}_{\text{QCD}} = \mbox{Re} \left[\frac{\lambda^{1/2}
    (M_{A_i}^2, m_{\tilde{t}_1}^2,m_{\tilde{t}_2}^2 )}{4\pi M_{A_i}^3}
  \left(\sum_{j=1}^2 {\bf Z}^*_{ij} G_{A_j}^{12\, *}\right) \,
  \frac{\alpha_s}{\pi} \, \left( \sum_{k=1}^2 
  {\bf Z}_{ik} \Delta^{\text{QCD}}_{A_k} \right) \right]
 \;, \label{maineq:QCD}
\eeq
where again the ${\bf Z}$ factors appear to ensure the on-shell
properties of the external loop-corrected Higgs field. The
$\Delta^{\text{QCD}}_{A_k}$ are given by the sum of the virtual, real 
and counterterm contributions $\Delta^V_{A_k}$, $\Delta^R_{A_k}$ and
$\Delta^{\text{CT}}_{A_k}$, respectively, 
\beq
\Delta^{\text{QCD}}_{A_k} = \Delta^V_{A_k} + \Delta^{\text{CT}}_{A_k} +
\Delta^R_{A_k} \;.
\eeq
From now on a factor $C_F \alpha_s/(4\pi)$ with $C_F = 4/3$ is
factorized out and already included in Eq.~(\ref{maineq:QCD}). The
virtual corrections receive contributions $\Delta^g_{A_k}$ from 
the gluon exchange diagram, $\Delta^{\tilde{g}}_{A_k}$
from the gluino exchange diagram and $\Delta^{4\tilde{t}}$ from the
diagram involving the 4-squark vertex. They are given by
\beq
\Delta^g_{A_k} &=& G_{A_k}^{12} \left[ B_0 (m_{\tilde{t}_1}^2; \zeta,
  m_{\tilde{t}_1}) +  B_0 (m_{\tilde{t}_2}^2; \zeta,m_{\tilde{t}_2}) -
  B_0 (M_{A_k}^2; m_{\tilde{t}_1}, m_{\tilde{t}_2}) \right. \nonumber \\
&& \left. + 2 (m_{\tilde{t}_1}^2 + m_{\tilde{t}_2}^2 - M_{A_k}^2 ) C_0 (\msta^2,
  M_{A_k}^2, \mstb^2; \zeta, \msta, \mstb) \right] 
\eeq
for the gluon exchange,
\beq
\Delta^{\tilde{g}}_{A_k} &=& \frac{g m_t {\cal R}^P_{k1}}{M_W \tan\beta}
\left[ \left(m_{\tilde{g}}-m_t \sin (2\theta_{\tilde{t}}) \right) B_0
  (\msta^2; m_{\tilde{g}}, m_t) + \left( m_{\tilde{g}} + m_t \sin
    (2\theta_{\tilde{t}}) \right) B_0(\mstb^2; m_{\tilde{g}},m_t)
\right. \nonumber \\
&&\left.
- \left( \sin (2\theta_{\tilde{t}}) m_t (\msta^2-\mstb^2) +
  m_{\tilde{g}} \mak^2 \right) C_0 (\msta^2,\mak^2,\mstb^2;
m_{\tilde{g}}, m_t, m_t) \right] 
\eeq
for the gluino exchange, and
\beq
\Delta^{4\tilde{t}}_{A_k} = G_{A_k}^{12} \, B_0 (\mak^2; \msta,\mstb) 
\eeq
for the 4-squark vertex diagram, where $B_0$ and $C_0$ are the
Passarino-Veltman scalar two- and three-point functions
\cite{loopintegs}, {\it cf.}~App.~\ref{app:scalardef} for their definitions. \s

\begin{figure}[t]
\begin{center}
\includegraphics[width=1.\textwidth]{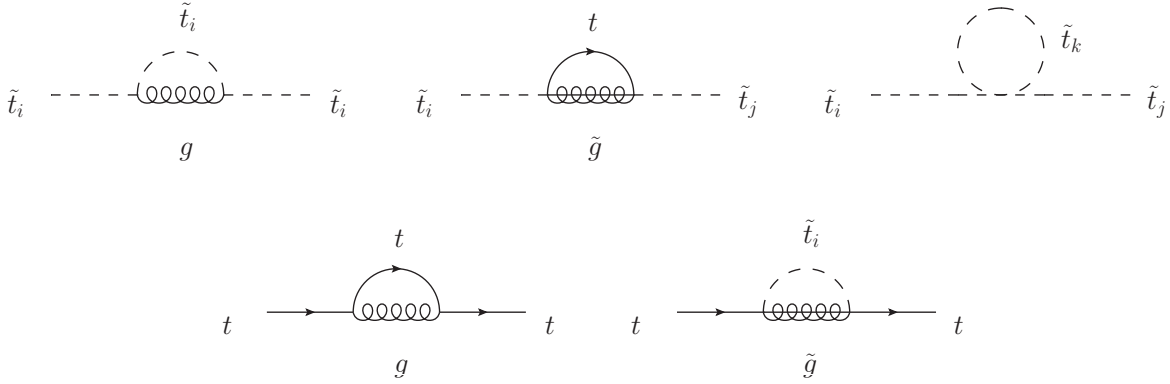}
\end{center}
\caption{Diagrams contributing at NLO SUSY-QCD to the squark self-energies
  (upper row) and to the quark self-energies (lower
  row). \label{fig:selfenergiesalphas}}  
\end{figure}

The counterterm corrections consist of the renormalization of the
external squark wave functions $Z_{\tilde{t}_{j} \tilde{t}_{j}}$ ($j=1,2$)
and the renormalization of the $A_k \tilde{t}_1 \tilde{t}_2$
interaction vertex. Note that the wave function renormalization of the
pseudoscalar does not contribute at order ${\cal O}(\alpha_s)$ and
hence to the SUSY QCD corrections. The parameters $\lambda$, $v$, $v_s$,
$\tan\beta$, $M_W$, $M_Z$ and $e$ are not
renormalized by the strong interaction, so that the counterterm reads 
\beq
\Delta_{A_k}^{\text{CT}} = \frac{1}{2} G_{A_k}^{12} \left( \delta
  Z_{\tilde{t}_1 \tilde{t}_1} + \delta Z_{\tilde{t}_2 \tilde{t}_2} \right) + \frac{\partial
  G_{A_k}^{12}}{\partial m_t} \delta m_t + \frac{\partial
  G_{A_k}^{12}}{\partial A_t} \delta A_t  \;,
\eeq
leading to
\beq
\Delta_{A_k}^{\text{CT}} = \frac{1}{2} G_{A_k}^{12} \left( \delta
  Z_{\tilde{t}_1 \tilde{t}_1} + \delta Z_{\tilde{t}_2 \tilde{t}_2} \right) + G_{A_k}^{12}
\frac{\delta m_t}{m_t} + \frac{g m_t}{2 M_W} \, \frac{{\cal
    R}^P_{k1}}{\tan\beta} \, \delta A_t \;. 
\eeq
The counterterm $\delta A_t$ of the trilinear coupling is given by the quark and
squark mass counter\-terms, $\delta m_t$ and $\delta
m_{\tilde{t}_{1,2}}$, and the mixing angle counterterm $\delta
\theta_{\tilde{t}}$,
\beq
\hspace{-2mm}\delta A_t = \frac{1}{2 m_t} \!\left[ (\msta^2-\mstb^2) \!\left( 2 \cos (2
    \theta_{\tilde{t}}) \delta \theta_{\tilde{t}} - \sin (2
    \theta_{\tilde{t}} ) \frac{\delta m_t}{m_t} \right)\! + 2 \sin
  (2 \theta_{\tilde{t}}) (\msta \delta \msta - \mstb \delta \mstb )
\right]\! . 
\eeq
In the (s)quark sector we adopt OS
renormalization with the quark and squark masses defined as the poles
of their respective propagators and the squark wave function
renormalization constants defined such that the residues of the poles
are equal to one. Defining the following structure for the quark
self-energy, where ${\cal P}_{L,R}$ denote, respectively, the left-
and right-chiral projector,
\beq
\Sigma_t (p^2) \equiv \slash{\!\!\!p} \Sigma^L_t (p^2) {\cal P}_L +
\slash{\!\!\!p} \Sigma^R_t (p^2) {\cal P}_R + m_t \Sigma^{Ls} (p^2) {\cal
  P}_L + m_t \Sigma^{Rs} (p^2) {\cal P}_R \;, 
\eeq
we have for the top quark mass counterterm
\beq
\delta m_t &=& \frac{1}{2} \mbox{Re} \left( m_t \Sigma_t^L (m_t^2) +
  m_t \Sigma_t^R (m_t^2) + \Sigma_t^{Ls} (m_t^2) + \Sigma_t^{Rs}
  (m_t^2) \right) \label{eq:delmt} \;.
\eeq
The squark mass and wave function counterterms are given by ($j=1,2$)
\beq
\delta m_{\tilde{t}_j}^2 &=& \mbox{Re} \Sigma_{\tilde{t}_j \tilde{t}_j}
(m_{\tilde{t}_j}^2) \label{eq:delmst} \\ 
\delta Z_{\tilde{t}_j \tilde{t}_j} &=& - \mbox{Re} \left. \frac{\partial
  \Sigma_{\tilde{t}_j \tilde{t}_j} (p^2)}{\partial
  p^2}\right|_{p^2=m_{\tilde{t}_j}^2} \;. \label{eq:delzst}
\eeq 
In Eq.~(\ref{eq:delmst}) the $\Sigma_{\tilde{t}_j \tilde{t}_j}$ denote
the diagonal parts of the squark self-energies. The diagrams
contributing at order ${\cal O}(\alpha_s)$ to the squark and quark 
self-energies are depicted in Fig.~\ref{fig:selfenergiesalphas}
(upper) and (lower), respectively. \s

The mass counterterms read
\beq
\frac{\delta m_t}{m_t} &=& - \mbox{Re} \Big[ 2 B_1 (m_t^2; m_t,
  \zeta) + B_1  (m_t^2; m_{\tilde{g}}, \msta) + B_1 (m_t^2;
  m_{\tilde{g}}, \mstb) + 4 B_0 (m_t^2; m_t, \zeta) \nonumber\\
&& + \sin (2\theta_{\tilde{t}}) \frac{m_{\tilde{g}}}{m_t}
  \left( B_0 (m_t^2; m_{\tilde{g}}, \msta) - B_0 (m_t^2;
    m_{\tilde{g}}, \mstb) \right) \Big] \\
\msta \delta \msta &=& \mbox{Re} \Big[2 m_t m_{\tilde{g}} \sin (2
\theta_{\tilde{t}} ) B_0 (\msta^2; m_t, m_{\tilde{g}}) +
\frac{1}{2}\left[ \left(1+\cos^2  (2\theta_{\tilde{t}}) \right) A_0
  (\msta) \right. \nonumber \\
&& \left. + \sin^2
  (2\theta_{\tilde{t}}) A_0 (\mstb) \right] - 2 \msta^2 B_0 (\msta^2;
\msta, \zeta) - A_0 (m_{\tilde{g}}) - A_0 (m_t) \nonumber \\
&& + (\msta^2-m_t^2-m_{\tilde{g}}^2) B_0 (\msta^2; m_t,
m_{\tilde{g}}) \Big]  \\
\mstb \delta \mstb &=& \mbox{Re} \Big[ -2 m_t m_{\tilde{g}} \sin (2
\theta_{\tilde{t}} ) B_0 (\mstb^2; m_t, m_{\tilde{g}}) +
\frac{1}{2}\left[ \left(1+\cos^2 (2\theta_{\tilde{t}}) \right) A_0
  (\mstb) \right. \nonumber \\
&& \left. + \sin^2
  (2\theta_{\tilde{t}}) A_0 (\msta) \right] - 2 \mstb^2 B_0 (\mstb^2;
\mstb, \zeta) - A_0 (m_{\tilde{g}}) - A_0 (m_t) \nonumber \\
&& + (\mstb^2-m_t^2-m_{\tilde{g}}^2) B_0 (\mstb^2; m_t,
m_{\tilde{g}}) \Big]  \;,
\eeq
where $B_1$ is the coefficient of the two-point tensor integral of rank
one and $A_0$ denotes the scalar one-point function, {\it
  cf.}~App.~\ref{app:scalardef}. 
The wave function corrections can be cast into the form
\beq
\delta Z_{\tilde{t}_1 \tilde{t}_1} &=& \mbox{Re} \Big[-4 m_t
m_{\tilde{g}} \sin (2\theta_{\tilde{t}}) B_0^\prime (\msta^2; m_t,
m_{\tilde{g}}) + 2 (m_{\tilde{g}}^2 + m_t^2 - \msta^2) B_0^\prime
(\msta^2; m_t, m_{\tilde{g}}) \nonumber \\
&& + 2 B_0 (\msta^2; \msta, \zeta) - 2 B_0 (\msta^2; m_t,
m_{\tilde{g}} ) + 4 \msta^2 B_0^\prime (\msta^2; \msta, \zeta) \Big]
\\
\delta Z_{\tilde{t}_2 \tilde{t}_2} &=& \mbox{Re} \Big[4 m_t
m_{\tilde{g}} \sin (2\theta_{\tilde{t}}) B_0^\prime (\mstb^2; m_t,
m_{\tilde{g}}) + 2 (m_{\tilde{g}}^2 + m_t^2 - \mstb^2) B_0^\prime
(\mstb^2; m_t, m_{\tilde{g}}) \nonumber \\
&& + 2 B_0 (\mstb^2; \mstb, \zeta) - 2 B_0 (\mstb^2; m_t,
m_{\tilde{g}} ) + 4 \mstb^2 B_0^\prime (\mstb^2; \mstb, \zeta) \Big] \;.
\eeq
Here $B_0^\prime (k^2; m_1^2, m_2^2)$ denotes the derivative with respect to
$k^2$. The mixing angle counterterm is renormalized as
\beq
\delta \theta_{\tilde{t}} = \frac{1}{2} (\delta Z_{\tilde{t}_1
  \tilde{t}_2} - \delta Z_{\tilde{t}_2 \tilde{t}_1}) =
\frac{1}{2(\msta^2-\mstb^2)} \mbox{Re} \big[\Sigma_{\tilde{t}_1 \tilde{t}_2}
(\mstb^2) + \Sigma_{\tilde{t}_2 \tilde{t}_1} (\msta^2)\big] \;, \label{eq:deltatheta}
\eeq
where $\Sigma_{\tilde{t}_i \tilde{t}_j}$ denotes the respective squark
self-energies, so that 
\beq
\delta \theta_{\tilde{t}} &=& \frac{1}{\msta^2-\mstb^2} \mbox{Re}
\left[ \sin   (2\theta_{\tilde{t}}) \cos (2\theta_{\tilde{t}}) \left(
    A_0 (\mstb^2) - A_0 (\msta^2) \right) \right. \nonumber \\
&& \left. + 2 m_t m_{\tilde{g}} \cos
  (2\theta_{\tilde{t}}) \left( B_0 (\mstb^2; m_t, m_{\tilde{g}}) + B_0
  (\msta^2; m_t, m_{\tilde{g}} ) \right) \right] \;.
\eeq
The real corrections finally in terms of dilogarithms read
\beq
\Delta_{A_k}^R &=& \frac{2 G_{A_k}^{12}}{\lambda^{1/2}} \Big[
(\mak^2-\msta^2-\mstb^2) \Big( -2 \log 
    \beta_0 \log \frac{\zeta \mak \msta \mstb}{\lambda} + 2 \log^2
    \beta_0 \nonumber \\
&&  - \log^2 \beta_1 - \log^2 \beta_2 + 2 \mbox{Li}_2
(1-\beta_0^2) - \mbox{Li}_2 (1-\beta_1^2) - \mbox{Li}_2 (1-\beta_2^2) \Big)
\nonumber \\
&& + 2 \lambda^{1/2} \log \frac{\zeta \mak \msta \mstb}{\lambda} + 4
\lambda^{1/2} + (2\mak^2+\msta^2+\mstb^2) \log \beta_0 \nonumber \\
&& + (\mak^2+2\mstb^2) \log \beta_2 + (\mak^2+2\msta^2) \log \beta_1 \Big]
\;,  \label{eq:realcorr}
\eeq
where in the two-body phase space function $\lambda (M_{A_k}^2,
  \msta^2, \mstb^2)$ we have neglected the arguments for better
  readability. We have furthermore introduced
\beq
\beta_0 &=& \frac{\mak^2-\msta^2-\mstb^2+\lambda^{1/2}}{2 \msta \mstb}
\nonumber \\
\beta_1 &=& \frac{\mak^2-\msta^2+\mstb^2-\lambda^{1/2}}{2 \mak \mstb} \\
\beta_2 &=& \frac{\mak^2+\msta^2-\mstb^2-\lambda^{1/2}}{2 \mak \msta}
\nonumber \;.
\eeq

\section{The One-Loop Electroweak Corrections \label{sec:ewcorr}}
The NLO electroweak corrections consist of the virtual corrections to
the vertex and the counterterm contributions to cancel the UV
divergences. Besides the top and stop fields and the parameters
specified below, again the Higgs field needs to be renormalized, 
as for the external Higgs field we use the two-loop corrected
Higgs boson mass at order ${\cal O}(\alpha_t \alpha_s)$, including the
full EW corrections at one-loop order. In the loops,
however, the tree-level masses for the Higgs bosons have to be used so
that the UV divergences are canceled properly. \s

\begin{figure}[t]
\begin{center}
\includegraphics[width=0.9\textwidth]{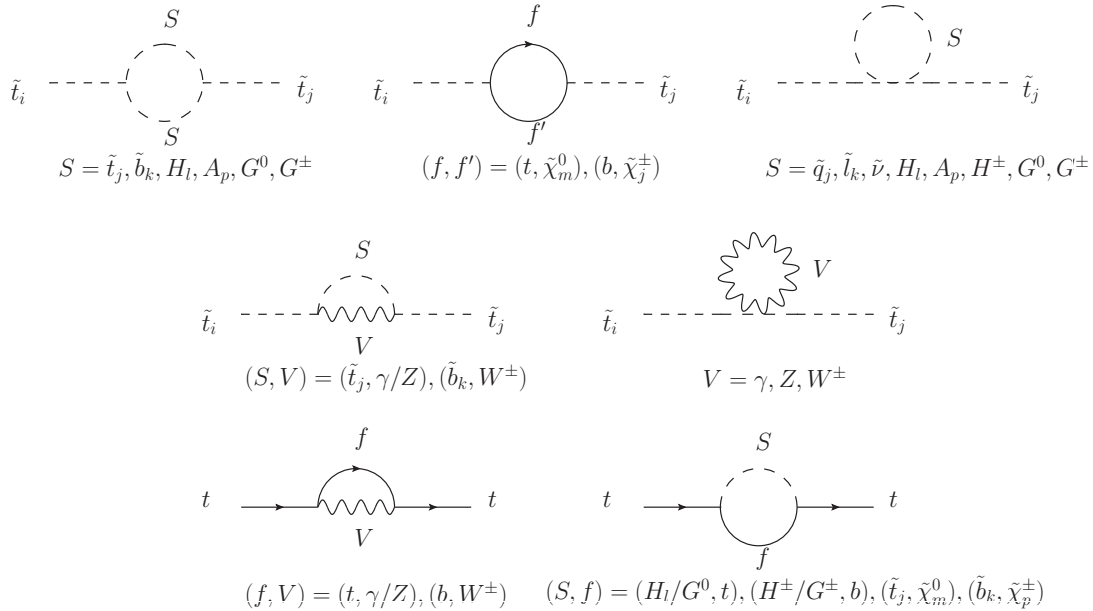}
\end{center}
\caption{Diagrams contributing to the electroweak squark self-energies
  (first two rows) and quark self-energies (last row) with $i,j,k,p=1,2$,
  $l=1,2,3$ and $m=1,...,5$. \label{fig:selfenergiesew}}
\end{figure}

For the NLO EW corrections, in addition the mixings 
\beq
\delta M^{G,Z}_{\text{mix},i} \equiv \delta M^{G,Z}_{\text{mix},i} (A_i \to
\tilde{t}_1 \tilde{t}_2) 
\eeq
of the decaying CP-odd Higgs boson $A_i$ with the $Z$
boson and the Goldstone boson $G$ have to be included. The matrix
element for the EW corrected pseudoscalar Higgs decay into stops hence reads
\beq
{\cal M} (A_i \to \tilde{t}_1 \tilde{t}_2) = \sum_{j=1}^2 {\bf
  Z}_{ij} \left( - G_{A_j}^{12} + \Delta^{\text{EW}}_{A_j} \right) +
\delta M^{G,Z}_{\text{mix},i} \;, 
\eeq 
where $\Delta^{\text{EW}}_{A_j}$ represents the sum of the 1-particle
irreducible (1PI) diagrams $\Delta^{V,\text{EW}}_{A_j}$ contributing to the EW virtual
corrections of the vertex and of the counterterms
$\Delta^{\text{CT},\text{EW}}_{A_j}$, 
\beq
\Delta^{\text{EW}}_{A_j} = \Delta^{V,\text{EW}}_{A_j} +
\Delta^{\text{CT},\text{EW}}_{A_j} \;. 
\eeq
Due to massless photons in the loops we
also encounter IR divergences. These are canceled by adding the real
corrections $\Delta^{R,\text{EW}}$,
where a photon is radiated off the final state stop lines. We hence
have for the EW corrected decay amplitude
\beq
\Gamma^{\text{NLO}}_{\text{EW}} (A_i \to \tilde{t}_1 \tilde{t}_2) =
\frac{3\lambda^{1/2}}{8\pi M_{A_i}^3} && \hspace*{-0.5cm}
\left[\; \left|\sum_{j=1}^2
  {\bf Z}_{ij} G_{A_j}^{12} \right|^2 + \left(\sum_{j=1}^2
  {\bf Z}^*_{ij} {G_{A_j}^{12}}^*\right) \left( \sum_{k=1}^2
  {\bf Z}_{ik} \Delta^{R,\text{EW}}_{A_k} \right) 
\right. \nonumber \\
\label{eq:nloewgam} 
&& \hspace*{-0.5cm} \left. - 2 \mbox{Re} \left( \sum_{j=1}^2
    {\bf Z}^*_{ij} {G_{A_j}^{12}}^* \left(\sum_{k=1}^2 {\bf Z}_{ik}
      \Delta^{\text{EW}}_{A_k} + \delta M^{G,Z}_{\text{mix,i}} \right)
  \right) \right]
\;.
\eeq
Again we have dropped the arguments in the two-body phase space
function $\lambda$. 
\begin{figure}[t]
\begin{center}
\includegraphics[width=0.9\textwidth]{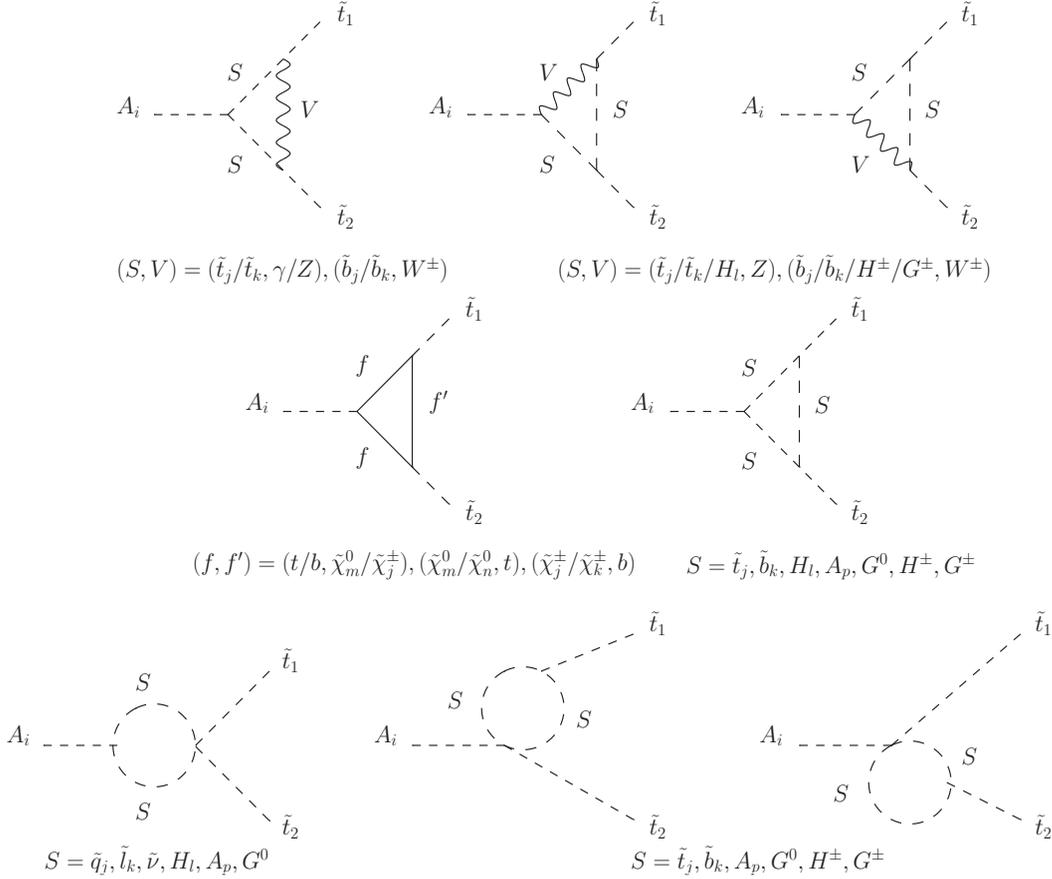}
\end{center}
\caption{Generic diagrams contributing to the electroweak corrections
  of the decay $\Gamma (A_i \to \tilde{t}_1
  \tilde{t}_2)$ with $i,j,k,p=1,2$, $l=1,2,3$ and $m,n=1,...,5$. \label{fig:triangleew}} 
\end{figure}
In the following we will discuss the individual contributions. The
virtual corrections consist of the 1PI 
diagrams given by the triangle diagrams with scalars,
fermions and gauge bosons in the loops, as shown in the first two rows
of Fig.~\ref{fig:triangleew}, and of the diagrams involving
four-particle vertices, {\it cf.}~Fig.~\ref{fig:triangleew} (last
row). For better readability, for the scalars $S$ appearing in the
loops we only listed the particle types but not the combination of
scalars that are allowed by the theory for the various vertices. Let
us remark, however, that in the four-particle vertices the
scalar-Higgs$-$Goldstone-boson$-$2-stops coupling
$H_l-G-\tilde{t}_1-\tilde{t}_2$ and the 
scalar$-$pseudoscalar$-$2-stops coupling $H_l-A_p-\tilde{t}_1-\tilde{t}_2$
($l=1,2,3$, $p=1,2$) are new compared to the MSSM. The former is due
to the singlet admixture in $H_l$, the latter is proportional to the
NMSSM specific coupling $\lambda$.  
The diagrams have been generated with the Mathematica package {\tt FeynArts} 3.6
\cite{feynarts} and evaluated with {\tt FormCalc} 7.3 \cite{formcalc}
in two independent calculations. The integrals have been computed with
{\tt LoopTools} 2.7\cite{formcalc}. The results of both calculations agree and have
been cross-checked against a third calculation, that did not use any
of the tools to evaluate and simplify the amplitudes, and which takes
the loop functions from {\tt HDECAY}~\cite{hdecay,susyhit} and {\tt
  SDECAY}~\cite{sdecay}. The UV divergences encountered in the
computation of the virtual 
corrections are canceled by the counterterms that are the sum of the
stop wave function corrections and of the counterterm renormalizing the
$A_j\tilde{t}_1\tilde{t}_2$ interaction,
\beq
\Delta_{A_j}^{\text{CT,EW}} = \Delta_{A_j}^{\text{CT},w} + \Delta_{A_j}^{\text{CT},v} 
\eeq
Because of the antisymmetric $G_{A_j}^{12}$ coupling the stop wave
function corrections are given by
\beq
\Delta_{A_j}^{\text{CT},w} = -\frac{G_{A_j}^{12}}{2} (\delta
Z_{\tilde{t}_1 \tilde{t}_1} + \delta Z_{\tilde{t}_2 \tilde{t}_2}) \;.
\eeq
The stops are renormalized on-shell, with the renormalization
conditions given in Eqs.~(\ref{eq:delmst}) and (\ref{eq:delzst}). The
diagrams, that contribute to the here required electroweak self-energies
are displayed in Fig.~\ref{fig:selfenergiesew} (upper two rows). \s

For the one-loop EW corrections the vertex counterterm reads
\beq
\Delta_{A_j}^{\text{CT},v} &=& -G^{12}_{A_j} \left( \frac{\delta g}{g} +
  \frac{\delta m_t}{m_t} - \frac{\delta M_W}{M_W} \right) - \frac{g m_t}{2
M_W} \left[ {\cal R}^P_{j1} \left( \delta \mu + \left( \frac{\delta
      A_t}{A_t} - \frac{\delta \tan\beta}{\tan\beta} \right)
  \frac{A_t}{\tan\beta} \right) \right. \nonumber \\
&& \left. + \left( \frac{\delta \lambda}{\lambda} + \frac{\delta v}{v}
    - \frac{\delta \tan\beta}{\tan\beta} \right) \frac{\lambda
    v}{\sqrt{2} \tan\beta} {\cal R}^P_{j2} \right] \;.
\label{eq:vertexcounter}
\eeq
The individual counterterms are derived from the renormalization of
the input parameters. We follow Ref.~\cite{Ender:2011qh} and apply the
same renormalization scheme which mixes OS and $\overline{\mbox{DR}}$
conditions as defined there. For the vertex counterterms the relevant
input parameters are the $W$ and $Z$ boson masses $M_W$ and $M_Z$, the
electric charge $e$, $\tan\beta$, $\lambda$ and
$v_s$.\footnote{Additionally, for the renormalization of the Higgs
  sector in the computation of the higher order corrections to the
  Higgs boson masses, we have the tadpole parameters, the mass of the
  charged Higgs boson $M_{H^\pm}$, the NMSSM
  parameter $\kappa$ and the trilinear coupling $A_\kappa$, which need
  to be renormalized at loop level, {\it cf.}~\cite{Ender:2011qh}.}
The parameters that can be related to physical quantities are renormalized
OS, the remaining ones $\overline{\mbox{DR}}$. Together with the
OS-renormalized top/stop sector, we have the following set of
parameters to be renormalized, 
\beq
\underbrace{\tan\beta,\lambda,v_s,}_{\overline{\mbox{DR}} 
  \mbox{ scheme}}\underbrace{M_Z, M_W, e,m_t,m_{\tilde{t}_1},m_{\tilde{t}_2},\theta_{\tilde{t}}}_{\mbox{OS
    scheme}} \;. 
\eeq
The coupling $g$ and the VEV $v$ appearing in
Eq.~(\ref{eq:vertexcounter}) are given in terms of these parameters by
\beq
g = \frac{eM_Z}{\sqrt{M_Z^2-M_W^2}} \qquad \mbox{and} \qquad
v = \frac{2M_W}{e} \sqrt{1-\frac{M_W^2}{M_Z^2}} \;,
\eeq
from which their counterterms can be derived. 
The details of the renormalization of the counter\-terms for the first six input
parameters can be found in \cite{Ender:2011qh}, so that they are not
repeated here. The formulae for the OS renormalization of the top and
stop masses are given in Eqs.~(\ref{eq:delmt})-(\ref{eq:delzst}). The
squark and quark self-energies for the EW one-loop corrections are
depicted in Fig.~\ref{fig:selfenergiesew}.  
At EW one-loop order the counterterm for $A_t$ now reads
\beq
\delta A_t &=& \frac{\mu}{\tan\beta} \left( \frac{\delta \mu}{\mu} -
  \frac{\delta \tan\beta}{\tan\beta} \right) + \frac{1}{2 m_t} \Big[
(m_{\tilde{t}_1}^2 - m_{\tilde{t}_2}^2 ) \left( 2 \delta
  \theta_{\tilde{t}} \cos (2 \theta_{\tilde{t}}) - \sin (2
  \theta_{\tilde{t}}) \frac{\delta m_t}{m_t} \right) \nonumber \\
&& + 2 \sin (2\theta_{\tilde{t}}) (m_{\tilde{t}_1} \delta
m_{\tilde{t}_1} - m_{\tilde{t}_2} \delta m_{\tilde{t}_2} ) \Big] \;.
\eeq
The counterterm for the mixing angle $\theta_{\tilde{t}}$ is
renormalized as in Eq.~(\ref{eq:deltatheta}), however with the self-energies
given by the diagrams shown in Fig.~\ref{fig:selfenergiesew} (first 
two rows). \s

\begin{figure}[t]
\begin{center}
\includegraphics[width=0.7\textwidth]{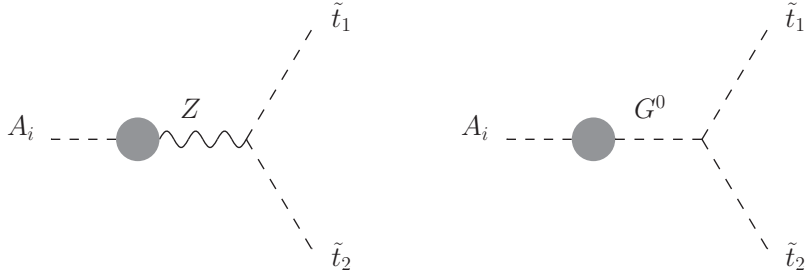}
\end{center}
\caption{Generic one-loop diagrams contributing to the mixing $\delta
  M^{G,Z}_{\text{mix},i}$ of a pseudoscalar $A_i$ with the $Z$ and Goldstone
  boson. \label{fig:agoldstonemix}}  
\end{figure}

The diagrams for the contributions to the electroweak corrections
stemming from the mixings of the pseudoscalar $A_i$ with the $Z$ boson
and the Goldstone boson, $\delta M^{G,Z}_{\text{mix},i}$, are shown in
Fig.~\ref{fig:agoldstonemix}.  
Using the Slavnov-Taylor identity \cite{dabelstein}, on can obtain the mixing
contributions through
\beq
\delta M^{G,Z}_{\text{mix},i} =  \frac{G_{G}^{12}}{M_Z} \, \hat{\Sigma}_{A_i Z}
\left( (M^{(0)}_{A_i})^2 \right) \;,
\eeq
where $\hat{\Sigma}_{A_i Z}$ denotes the renormalized $A_i$-$Z$ mixing
self-energy and $G_{G}^{12}$ is the Goldstone boson coupling
to the stops,
\beq
G_{G}^{12} = -\frac{g m_t}{2 M_W} \left( A_t -
  \frac{\mu}{\tan\beta}\right) \;.
\eeq
Note, in particular that the external momenta have to be set to the
tree-level mass $M^{(0)}_{A_i}$ in order to ensure to maintain gauge
invariance. \s

The last piece which is missing in the decay, in order to also get an
IR finite result, is the real corrections term
$\Delta_{A_k}^{R,\text{EW}}$. This is the same as for the QCD corrections,
but with the gluon replaced by the photon. In the formula for the QCD
corrections, Eq.~(\ref{eq:realcorr}), this means that the coupling and
color factors have to be replaced accordingly, {\it i.e.}
\beq
\Delta_{A_k}^{R,\text{EW}} =  \frac{e^2}{16 \pi^2}
\Delta_{A_k}^{R} \;,
\eeq 
with $\Delta_{A_k}^R$ given in Eq.~(\ref{eq:realcorr}). \s

The full NLO decay width including the SUSY-QCD and -EW corrections is
then given by 
\beq
\Gamma^{\text{NLO,}\, \text{full}} = \Gamma^{\text{NLO}}_{\text{EW}} +
\Gamma^{(1)}_{\text{QCD}} \;,
\eeq
where $\Gamma^{\text{NLO}}_{\text{EW}}$, defined in
Eq.~(\ref{eq:nloewgam}), includes the leading order decay width and
$\Gamma^{(1)}_{\text{QCD}}$ has been defined in Eq.~(\ref{eq:nloqcdgam}).

\section{Numerical Analysis \label{sec:numerical}}
For our numerical analysis we first perform a scan in the NMSSM
parameter space in order to find scenarios that are in accordance with
the LHC Higgs and SUSY data. The compatibility with the LHC Higgs data
has been checked by using the programs {\tt HiggsBounds}
\cite{higgsbounds} and {\tt HiggsSignals} \cite{higgssignals}. The
effective couplings of the NMSSM Higgs bosons, normalized to the 
corresponding SM values, as well as the masses, the widths and the branching
ratios of the NMSSM Higgs bosons, which are required as inputs for these
programs, have been obtained from the Fortran code {\tt NMSSMCALC}
\cite{Baglio:2013iia}. The loop induced
Higgs coupling to gluons normalized to the corresponding coupling of a
SM Higgs boson with same mass is obtained by taking the ratio of the
partial widths for the Higgs decays into gluons in the NMSSM and the
SM, respectively. These include the QCD corrections up to
next-to-next-to-next-to leading order in the limit of heavy quarks
\cite{Inami:1982xt,Djouadi:1991tka,Spira:1993bb,Spira:1995rr,Kramer:1996iq,Chetyrkin:1997iv,Chetyrkin:1997un,Schroder:2005hy,Chetyrkin:2005ia,Baikov:2006ch}
and squarks \cite{Dawson:1996xz,Djouadi:1996pb}, taken over from the
SM, respectively, MSSM case, while the EW are unknown for the SUSY
case and hence consistently neglected also in the SM decay width. 
The stop mass values have been chosen such that they are not excluded by present
ATLAS \cite{atlasstop1,atlasstop2,atlasstop3} and CMS
\cite{cmsstop1,cmsstop2} searches. The squark 
masses of the first two generation are heavy enough not to be in
conflict with LHC data. The SM input parameters that we use are 
\cite{Agashe:2014kda,Jegerlehner:2011mw}
\begin{align}
\alpha(M_Z)&=1/128.962 \,, &\alpha^{\overline{\mbox{MS}}}_s(M_Z)&=
0.1184 \,, &M_Z&=91.1876\, 
\gev\,, \label{eq:param1} \\ \non
M_W&=80.385\,\gev \,,   &m_t&=173.5\,\gev \,,
&m^{\MSb}_b(m_b^{\MSb})&=4.19\,\gev \;. 
\end{align} 
The running $\alpha_s^{\DRb}$ used in {\tt NMSSMCALC} is obtained by
converting the $\alpha^{\overline{\mbox{MS}}}_s$, that is evaluated with the SM
renormalization group equations at two-loop order, to the $\DRb$
scheme. The light quark masses, which have only a small
influence on the loop results, have been set to  
\beq
m_u=2.5\,\mev\; , \quad m_d=4.95\,\mev\; , \quad m_s=100\, \mev \quad
\mbox{and} \quad m_c=1.42\, \gev \;. \label{eq:param2}
\eeq
As renormalization scale $\mu_R$ we choose the SUSY scale $M_s$, which
we set
\beq
M_s = \sqrt{m_{\tilde Q_3}m_{\tilde t_R}} = \mu_R \;.
\eeq

\subsection{Pseudoscalar Higgs Boson Decays into Stop Pairs}
In this subsection we present the impact of the SUSY-QCD and -EW
corrections on the decay of a heavy pseudoscalar Higgs boson into a
pair of stop quarks. The parameter point, which we have chosen from the
set of parameter points that survive the LHC constraints, is given by
the soft SUSY breaking masses and trilinear couplings 
\beq
&&  m_{\tilde{u}_R,\tilde{c}_R} = 
m_{\tilde{d}_R,\tilde{s}_R} =
m_{\tilde{Q}_{1,2}}= m_{\tilde L_{1,2}} =m_{\tilde e_R,\tilde{\mu}_R} = 3\;\mbox{TeV}\, , \;  
m_{\tilde{t}_R}=536.43\,\gev \,,\; \non \\ \non
&&  m_{\tilde{Q}_3}=594.61\,\gev\,,\; m_{\tilde{b}_R}=1285\,\gev\,,\; 
m_{\tilde{L}_3}=255.53\,\gev\,,\; m_{\tilde{\tau}_R}=1499\,\gev\,,
 \\ 
&& A_t = 1418\,\gev\, ,\; A_{u,c} = 1435\,\gev\, ,\;
A_{d,s,b}=-66.68\,\gev\,,\; A_{e,\mu,\tau} = -91.76\,\gev\,,\;  \non \\ 
&& M_1 = 111.73\,\gev,\; M_2= 395.86\,\gev\,,\; M_3=1370\,\gev 
\label{eq:param3}
\eeq
and NMSSM specific input parameters 
\beq
&& \lambda = 0.629 \;, \quad \kappa = 0.223 \; , \quad A_\kappa = -543.53\,\gev\;,\quad 
\mu_{\text{eff}} = 452.61\,\gev \;, \non \\ 
&& \tan\beta = 1.969 \;,\quad M_{H^\pm} = 1024\,\gev \;. \label{eq:param4}
\eeq
This results in the two-loop corrected mass $M_{A_2}$ of the heavy
pseudoscalar $A_2$ and the stop masses $m_{\tilde{t}_{1,2}}$,
\beq
M_{A_2}= 1012 \mbox{ GeV} \;, \quad m_{\tilde{t}_1} = 280.78 \mbox{ GeV} \quad
\mbox{and} \quad
m_{\tilde{t}_2} = 709.07 \mbox{ GeV} \;.
\eeq
We follow the SLHA format \cite{slha}, in which the parameters $\lambda, \kappa,
A_\kappa, \mu_{\text{eff}}, \tan\beta$ as well as the soft 
SUSY breaking masses and trilinear couplings are understood as $\DRb$
parameters at the scale $\mu_R = M_s$\footnote{For $\tan\beta$ this is
the case only, if it is read in from the block EXTPAR. Otherwise it is the 
$\overline{\mbox{DR}}$ parameter at the scale $M_Z$.}, whereas the
charged Higgs mass is an OS parameter. As input for our computation 
we take the soft SUSY breaking trilinear stop coupling $A_t$,
however, consistently as OS parameter. The conversion from the
$\overline{\mbox{DR}}$ to the OS scheme is done within
{\tt NMSSMCALC} and yields\footnote{Note, however, that the conversion is done
  through a counterterm that involves, as required for 
  the order ${\cal O}(\alpha_t \alpha_s)$ corrections computed in
  {\tt NMSSMCALC} order ${\cal O}(\alpha_s)$ corrections and no EW corrections.}
\beq
A_t^{\text{OS}} = 1435\,\gev \;. \label{eq:param5}
\eeq
The leading order width obtained in this scenario amounts to
\beq
\Gamma^{\text{LO}} (A_2 \to \tilde{t}_1 \tilde{t}_2) = 15.72\,\gev \;,
\eeq
where again we have summed over both charge conjugated stop pair final
states. The LO width differs by 2.7\% from the value obtained at tree
level with {\tt NMSSMCALC}, where the Fermi constant $G_F$ instead of
$\alpha$ is used as input parameter. \s
\begin{figure}[t!]
\begin{center}
\includegraphics[width=0.46\textwidth]{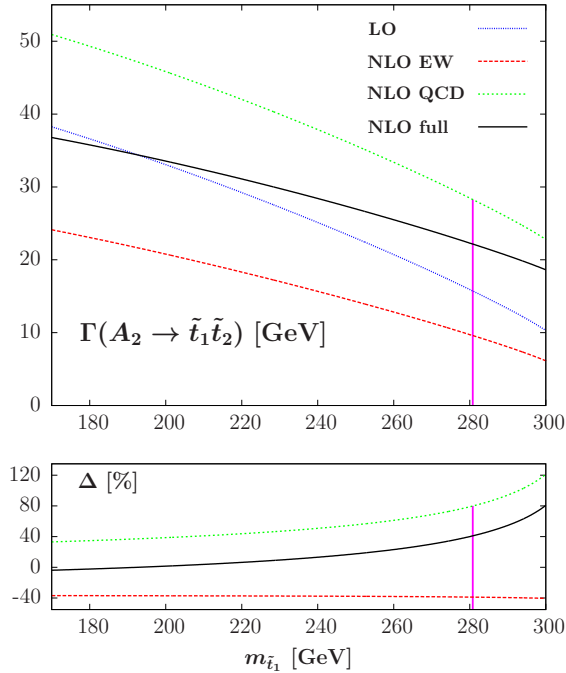}
\end{center}
\caption{Upper: The partial decay width $\Gamma(A_2 \to \tilde{t}_1
    \tilde{t}_2)$ as a function of $m_{\tilde{t}_1}$ 
   at LO (blue/lower dotted), including the NLO QCD (green/upper
  dotted), the NLO EW (red/dashed) and both the EW and QCD corrections 
  (black/full). Lower: The relative correction $\Delta = (\Gamma^{\text{NLO}}_X -
  \Gamma^{\text{LO}})/\Gamma^{\text{LO}}$ in per cent for $X=$QCD
  (green/dotted), EW (red/dashed) and the full NLO corrections
  (black/full). The pink line shows the position of the parameter
  point defined in Eqs.~(\ref{eq:param1})--(\ref{eq:param4}). \label{fig:nlowidthat}}  
\end{figure}

In Fig.~\ref{fig:nlowidthat} (upper) we show the partial decay width $\Gamma
(A_2 \to \tilde{t}_1 \tilde{t}_2)$ at LO, including the EW and the QCD
corrections, and the NLO width with both the QCD and EW corrections
taken into account, as a function of $m_{\tilde{t}_1}$, which is
varied around the parameter point defined in
Eqs.~(\ref{eq:param1})--(\ref{eq:param4}) with $m_{\tilde{t}_1} \approx
281$~GeV.\footnote{The variation of $m_{\tilde{t}_1}$ between 170 and
  300~GeV corresponds to a variation of $A_t^{\text{OS}}$ between 1371
  and 1721~GeV.} Note, that stops can still be rather light
\cite{atlasstop2,atlasstop3,cmsstop2,Muhlleitner:2011ww,Grober:2014aha}, 
down to about 240 GeV for arbitrary neutralino masses \cite{atlasstop3}
and even lower taking into account the actual $\tilde{t}_1$ branching
ratio \cite{Grober:2014aha}. The figure illustrates the effect of the
higher order corrections, although the thus obtained parameter 
configurations are not all in accordance with the applied constraints
anymore. The lower plot displays the relative corrections 
\beq
\Delta = \frac{\Gamma^{\text{NLO}}_X -
  \Gamma^{\text{LO}}}{\Gamma^{\text{LO}}} \;, \quad X = \text{QCD},
\text{ EW}, \text{ QCD+EW} 
\eeq 
in per cent. The plots show that both the QCD and the EW corrections
are significant and come with opposite sign. The QCD correction
increase the LO width by $\sim 40-120$\% in the investigated range,
depending on the value of $m_{\tilde{t}_1}$, whereas the EW
corrections are almost independent of $m_{\tilde{t}_1}$ and decrease
the cross section by 40\%. At $m_{\tilde{t}_1} =192$~GeV the QCD and
EW corrections are of same size and cancel each other. The full
corrections hence increase the width between $\sim 0-80$\%, {\it
  cf.}~Fig.~\ref{fig:nlowidthat} (lower). And for
our parameter point the total correction is
\beq
\Delta \Gamma^{\text{QCD+EW}} (A_2 \to \tilde{t}_1 \tilde{t}_2) = 41\% \;.
\eeq
This plot demonstrates that both the inclusion of the EW and the QCD corrections is
required in order to properly predict the decay width. \s

\begin{figure}[h!]
\begin{center}
\includegraphics[width=0.46\textwidth]{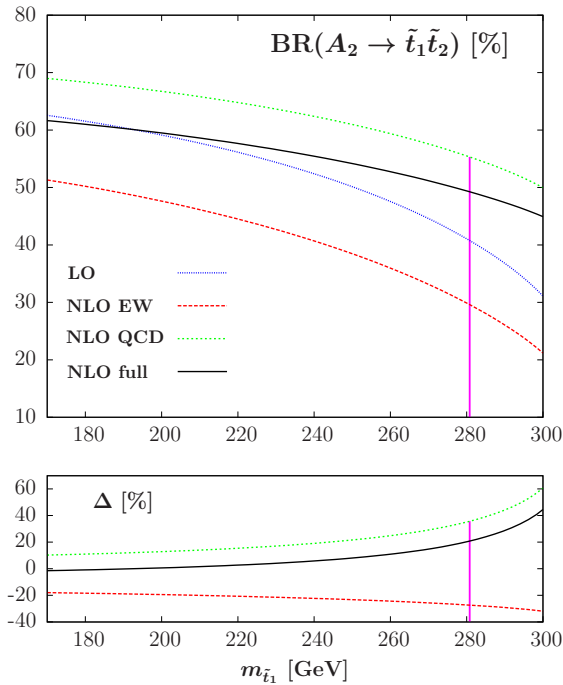}
\end{center}
\caption{Same as Fig.~\ref{fig:nlowidthat}, but for the branching
  ratios. \label{fig:branratios}}   
\end{figure}
In Fig.~\ref{fig:branratios} we show the branching ratios
corresponding to the widths of Fig.~\ref{fig:nlowidthat}. They have
been obtained by replacing in {\tt NMSSMCALC} the corresponding width
with our loop corrected width.\footnote{In {\tt NMSSMCALC} the SUSY
  QCD corrections to the decays into squarks, as derived from
  \cite{Accomando:2011jy}, are taken into account. These include 
  improvements in the decays into sbottoms, which are required in parts
  of the parameter space, that are not relevant for us. Furthermore, we
  include the EW corrections. We therefore consistently turned off the
  corrections implemented in {\tt NMSSMCALC} in the decays into
  squarks and included instead our corrections.} The 
branching ratio at LO of our investigated parameter point amounts to
\beq
\mbox{BR}^{\text{LO}} (A_2 \to \tilde{t}_1 \tilde{t}_2)= 40.8\% \;.
\eeq 
The net effect of the NLO EW and QCD corrections is an increase of the
branching ratio by
\beq
\Delta \mbox{BR}^{\text{QCD+EW}} (A_2 \to \tilde{t}_1 \tilde{t}_2) = 20.8\% \;.
\eeq
In the plot of Fig.~\ref{fig:branratios} we again vary
$m_{\tilde{t}_1}$ around the chosen parameter 
point, illustrated by the pink line in the plot. As can be read off
the lower plot the total NLO corrections increase the branching ratio
by up to a bit more than $40$\% in the shown $m_{\tilde{t}_1}$ range. Thus this
decay remains the most important one also after inclusion of the NLO
corrections. \s 

\begin{figure}[ht!]
\begin{center}
\includegraphics[width=0.46\textwidth]{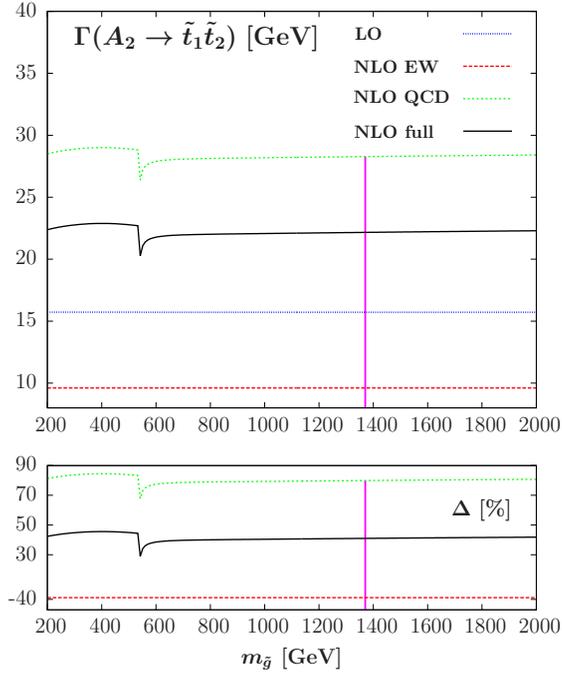}
\end{center}
\caption{Upper: The partial decay width $\Gamma(A_2 \to \tilde{t}_1
    \tilde{t}_2)$ as a function of $m_{\tilde{g}}$. The color and line
    style code is the same as in Fig.~\ref{fig:nlowidthat}. Lower: The
    relative correction $\Delta = (\Gamma^{\text{NLO}}_X - 
  \Gamma^{\text{LO}})/\Gamma^{\text{LO}}$ in per cent for $X=$QCD
  (green/dotted), EW (red/dashed) and the full NLO corrections
  (black/full). The pink line shows the position of the parameter
  point defined in Eqs.~(\ref{eq:param1})-(\ref{eq:param4}). \label{fig:mgluinodep}}  
\end{figure}

Figure~\ref{fig:mgluinodep} finally shows the dependence of the higher
order corrections on the gluino mass. The EW corrections of course do
not depend on $m_{\tilde{g}}$, while the QCD corrections show a very
mild dependence on the gluino mass, apart from the region around
$m_{\tilde{g}} \approx 535$~GeV. The kink that
appears here, arises in the $\tilde{t}_2$ self-energy at the threshold where
$m_{\tilde{t}_2} = m_{\tilde{g}} + m_t$. \s

\begin{figure}[ht!]
\begin{center}
\includegraphics[width=0.46\textwidth]{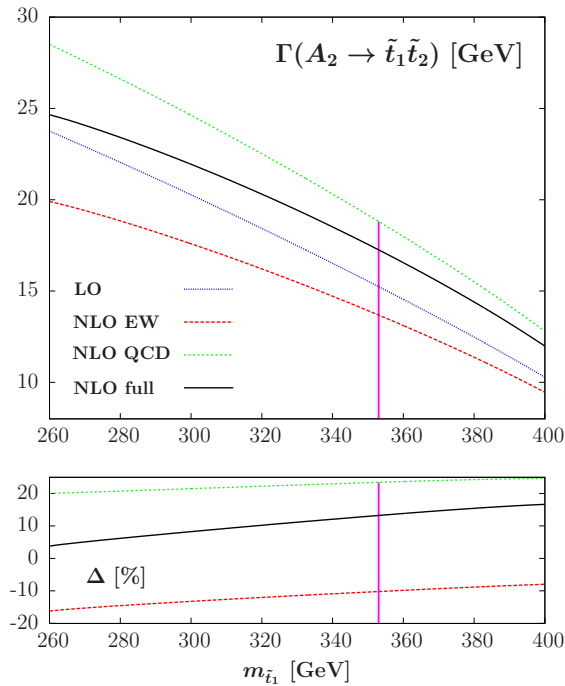}
\end{center}
\caption{Same as Fig.~\ref{fig:nlowidthat}, but now for
  the initial scenario (marked by the pink line in the plot) given by
  Eqs.~(\ref{eq:parama})--(\ref{eq:paramb}) 
\label{fig:scen2decay}}  
\end{figure}

The size of the higher order corrections sensitively depends on the
scenario. Thus we find for the scenario defined by
\beq
&&  m_{\tilde{u}_R,\tilde{c}_R} = 
m_{\tilde{d}_R,\tilde{s}_R} =
m_{\tilde{Q}_{1,2}}= m_{\tilde L_{1,2}} =m_{\tilde e_R,\tilde{\mu}_R} = 3\;\mbox{TeV}\, , \;  
m_{\tilde{t}_R}=714.25\,\gev \,,\; \non \\ \non
&&  m_{\tilde{Q}_3}=1035\,\gev\,,\; m_{\tilde{b}_R}=2776\,\gev\,,\; 
m_{\tilde{L}_3}=2156\,\gev\,,\; m_{\tilde{\tau}_R}=1755\,\gev\,,
 \\ 
&& A_t = 1246\,\gev\, ,\; A_{u,c} = 1347\,\gev\, ,\;
A_{d,s,b}=-1651\,\gev\,,\; A_{e,\mu,\tau} = 769.08\,\gev\,,\;  \non \\ 
&& M_1 = 460.61\,\gev,\; M_2= 381.55\,\gev\,,\; M_3=2296\,\gev 
\label{eq:parama}
\eeq
and 
\beq
&& \lambda = 0.552 \;, \quad \kappa = 0.030 \; , \quad A_\kappa = -173.51\,\gev\;,\quad 
\mu_{\text{eff}} = 446.80\,\gev \;, \non \\ 
&& \tan\beta = 3.005 \;,\quad M_{H^\pm} = 1460\,\gev
\;, \label{eq:paramb} 
\eeq
resulting in
\beq
M_{A_2}= 1461 \mbox{ GeV} \;, \quad m_{\tilde{t}_1} = 353.02 \mbox{ GeV} \quad
\mbox{and} \quad
m_{\tilde{t}_2} = 927.56 \mbox{ GeV} 
\eeq
and an LO decay width of
\beq
\Gamma^{\text{LO}} (A_2 \to \tilde{t}_1 \tilde{t}_2 ) = 15.24 \mbox{ GeV}
\eeq 
NLO QCD and EW corrections that amount to
$\Delta^{\text{QCD}} = 23.4$\% and $\Delta^{\text{EW}} = -10.2$\%,
respectively, resulting in a total correction of 
\beq
\Delta^{\text{QCD+EW}} (A_2 \to \tilde{t}_1 \tilde{t}_2) = 13.2\% \;.
\eeq
This can also be inferred from Fig.~\ref{fig:scen2decay} which
shows the NLO corrections to the decay widths as a function of
$m_{\tilde{t}_1}$. As demonstrated in the lower plot, the NLO QCD
corrections are of the order of 20-25\%, while the EW corrections
range between about -17\% and -8\%, leading to an overall increase of
the cross section between 3\% and 17\% due to the combined NLO corrections.

\subsection{Stop Decays into a Pseudoscalar}
Decays of the heavy stop into a pseudoscalar and the light stop, 
\beq
\tilde{t}_2 \to A_i + \tilde{t}_1\;, \qquad i=1,2 \;,
\eeq
can occur and become important when there is a large mass splitting
between the two stop mass eigenstates. Stop decays into
a SM-like Higgs boson final state have recently been discussed in
\cite{Ghosh:2013qga} and the production of NMSSM Higgs bosons in
squark cascade decays in
\cite{Ellwanger:2014hca,Chakraborty:2015xia}. The LO decay 
width is obtained from Eq.~(\ref{eq:treegam}) by adapting the
kinematic factor and dividing by the color factor 3 and the factor 2
due to the summation of the two charge conjugated stop pair final
states, 
\beq
\Gamma^{\text{LO}} (\tilde{t}_2 \to A_i \tilde{t}_1) =
\frac{\lambda^{1/2} (m_{\tilde{t}_2}^2,M_{A_i}^2,
  m_{\tilde{t}_1}^2)}{16 \pi m_{\tilde{t}_2}^3} 
\left| \sum_{j=1}^2 {\bf Z}_{ij} G_{A_j}^{12} \right|^2
 \;.
\label{eq:treegamstop}
\eeq
The NLO SUSY-QCD decay width can be written as
\beq
\Gamma^{\text{NLO}} (\tilde{t}_2 \to A_i \tilde{t}_1) =
\Gamma^{\text{LO}} (\tilde{t}_2 \to A_i \tilde{t}_1)  + 
\Gamma^{(1)}_{\text{QCD}} (\tilde{t}_2 \to A_i \tilde{t}_1) \;,
\eeq
where
\beq
\Gamma^{(1)}_{\text{QCD}} (\tilde{t}_2 \to A_i \tilde{t}_1)= \mbox{Re} \left[
\frac{\lambda^{1/2} (m_{\tilde{t}_2}^2,M_{A_i}^2,
  m_{\tilde{t}_1}^2)}{24\pi m_{\tilde{t}_2}^3} \, 
\left( \sum_{j=1}^2 {\bf Z}^*_{ij} {G_{A_j}^{12}}^* \right) \,
\frac{\alpha_s}{\pi} \, \left( \sum_{k=1}^2 {\bf Z}_{ik}
  \Delta^{\text{QCD}}_{k,\tilde{t}_2} \right) \right] \;,  
\label{maineq:QCDstop}
\eeq
with the correction factor
\beq
\Delta^{\text{QCD}}_{k,\tilde{t}_2} = \Delta^V_{k,\tilde{t}_2} +
\Delta^{\text{CT}}_{k,\tilde{t}_2} + \Delta^R_{k,\tilde{t}_2} \;.
\eeq
The virtual corrections\footnote{The diagrams are the same as in
  Fig.~\ref{fig:nlosusyqcd}, but with 
  $\tilde{t}_2$ in the initial and $A_i$ in the final state.}
and the counterterms are given by the same 
expressions as for the pseudoscalar decay presented in
section~\ref{sec:susyqcd}, {\it i.e.} 
\beq
\Delta^V_{k,\tilde{t}_2} = \Delta^V_{A_k} \qquad \mbox{and} \qquad
\Delta^{\text{CT}}_{k,\tilde{t}_2} = \Delta^{\text{CT}}_{A_k} \;.
\eeq
In the real corrections, though, the roles of $A_k$ and $\tilde{t}_2$
have to be interchanged,
\beq
\Delta^R_{k,\tilde{t}_2} = \Delta^R_{A_k} (M_{A_k}^2 \leftrightarrow
m_{\tilde{t}_2}^2) \;, \label{eq:realcorrst2}
\eeq
with $\Delta_{A_k}^R$ given in Eq.~(\ref{eq:realcorr}).
The NLO SUSY-EW corrections are composed of the same electroweak
virtual corrections to the vertex\footnote{In the Feynman diagrams of
  Fig.~\ref{fig:triangleew} simply the $\tilde{t}_2$ leg has to be crossed to the
  initial and the $A_i$ leg to the final state.}, the same
counterterms and the same mixing 
contributions of the pseudoscalar with the $Z$ and Goldstone boson,
$\Delta^{\text{EW}}_{A_k}$ and $\delta M^{G,Z}_{\text{mix},i}$, as in the
pseudoscalar decay. In the real corrections, however, again 
$M_{A_i}^2$ and $m_{\tilde{t}_2}^2$ have to be interchanged. We hence
have 
\beq
\Gamma^{\text{NLO}}_{\text{EW}} (\tilde{t}_2 \to A_i \tilde{t}_1) =
\frac{\lambda^{1/2}}{16\pi
  m_{\tilde{t}_2}^3} && \hspace*{-0.5cm} \left[\; \left|\sum_{j=1}^2 
  {\bf Z}_{ij} G_{A_j}^{12} \right|^2
+ \left( \sum_{j=1}^2 {\bf Z}_{ij}^*
    {G_{A_j}^{12}}^* \right) 
\left( \sum_{k=1}^2 {\bf Z}_{ik} \Delta^{R,\text{EW}}_{k,\tilde{t}_2}
\right) \right. \nonumber \\
&& \hspace*{-0.5cm} \left. - 2 \mbox{Re} \left( \sum_{j=1}^2
    {\bf Z}^*_{ij} {G_{A_j}^{12}}^* \left(\sum_{k=1}^2 {\bf Z}_{ik}
      \Delta^{\text{EW}}_{A_k} + \delta M^{G,Z}_{\text{mix},i} \right)
  \right) \right]
\;, \label{eq:ewcorrst2}
\eeq
with
\beq
\lambda^{1/2} \equiv \lambda^{1/2} (m_{\tilde{t}_2}^2,M_{A_i}^2,m_{\tilde{t}_1}^2)
\eeq
and
\beq
\Delta_{k,\tilde{t}_2}^{R,\text{EW}} =  \frac{e^2}{16 \pi^2}
\Delta^R_{A_k} (M_{A_k}^2 \leftrightarrow
m_{\tilde{t}_2}^2)
\eeq 
and $\Delta^{\text{EW}}_{A_k}$ and $\delta M^{G,Z}_{\text{mix},i}$ given in
section~\ref{sec:ewcorr}. The full NLO decay width including SUSY-QCD and
-EW corrections is given by 
\beq
\Gamma^{\text{NLO, full}} = \Gamma^{\text{NLO}}_{\text{EW}} +
\Gamma^{(1)}_{\text{QCD}} \;,
\eeq
with $\Gamma^{(1)}_{\text{QCD}}$ given by
Eqs.~(\ref{maineq:QCDstop})--(\ref{eq:realcorrst2}) and
$\Gamma^{\text{NLO}}_{\text{EW}}$ by Eq.~(\ref{eq:ewcorrst2}).\s

\begin{figure}[t!]
\begin{center}
\includegraphics[width=0.46\textwidth]{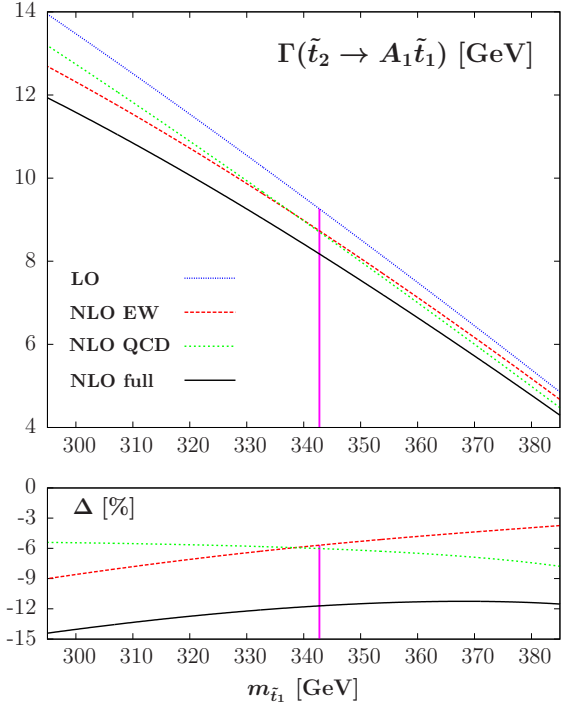}
\end{center}
\caption{The decay width $\Gamma (\tilde{t}_2 \to A_1 \tilde{t}_1)$ as
  a function of $m_{\tilde{t}_1}$ at LO
  (blue/upper dotted), including the NLO QCD (green/lower 
  dotted), the NLO EW (red/dashed) and both the EW and QCD corrections 
  (black/full). Lower: The relative correction $\Delta = (\Gamma^{\text{NLO}}_X -
  \Gamma^{\text{LO}})/\Gamma^{\text{LO}}$ in per cent for $X=$QCD
  (green/dotted), EW (red/dashed) and the full NLO corrections
  (black/full). The pink line shows the position of the parameter
  point defined in Eqs.~(\ref{eq:scen2param1})-(\ref{eq:scen2param2}). \label{fig:stopdecays}}  
\end{figure}
In order to show the impact of the SUSY-EW and -QCD corrections on the
stop decay width we chose the following parameter set, which leads to
an NMSSM Higgs and SUSY spectrum in accordance with the LHC data:
\beq
&&  m_{\tilde{u}_R,\tilde{c}_R} = 
m_{\tilde{d}_R,\tilde{s}_R} =
m_{\tilde{Q}_{1,2}}= m_{\tilde L_{1,2}} =m_{\tilde e_R,\tilde{\mu}_R} = 3\;\mbox{TeV}\, , \;  
m_{\tilde{t}_R}=748.07\,\gev \,,\; \non \\ \non
&&  m_{\tilde{Q}_3}=1259\,\gev\,,\; m_{\tilde{b}_R}=1709\,\gev\,,\; 
m_{\tilde{L}_3}=1637\,\gev\,,\; m_{\tilde{\tau}_R}=1618\,\gev\,,
 \\ 
&& A_t = 1589\,\gev\, ,\; A_{u,c} = 1675\,\gev\, ,\;
A_{d,s,b}=-669.04\,\gev\,,\; A_{e,\mu,\tau} = 179.93\,\gev\,,\;  \non \\ 
&& M_1 = 645.51\,\gev,\;  M_2= 272.11\,\gev\,,\; M_3=2511\,\gev 
 \;, \label{eq:scen2param1}
\eeq
and the NMSSM specific input parameters 
\beq
&& \lambda = 0.588 \;, \quad \kappa = 0.378 \; , \quad A_\kappa = -675.73\,\gev\;,\quad 
\mu_{\text{eff}} = 385.90\,\gev \;, \non \\ 
&& \tan\beta = 1.529 \;,\quad M_{H^\pm} = 639.08\,\gev
\;. \label{eq:scen2param2} 
\eeq
This results in the two-loop corrected mass $M_{A_1}$ of the lighter 
pseudoscalar $A_1$ and the stop masses $m_{\tilde{t}_{1,2}}$,
\beq
M_{A_1}= 637.40 \mbox{ GeV} \;, \quad m_{\tilde{t}_1} = 342.76 \mbox{
  GeV} \quad \mbox{and} \quad m_{\tilde{t}_2} = 1153 \mbox{ GeV} \;.
\eeq
And for the stop soft SUSY breaking trilinear coupling in the OS
scheme we get 
\beq
A_t^{\text{OS}} = 1675 \mbox{ GeV} \;.
\eeq

Figure~\ref{fig:stopdecays} (upper) shows the decay width $\Gamma (\tilde{t}_2
\to A_1 \tilde{t}_1)$ at LO, including the NLO EW, the NLO QCD and
both NLO corrections, as a function of $m_{\tilde{t}_1}$, varied
around the chosen parameter point, 
marked by the pink line in the plots. The LO decay width reaches
\beq
\Gamma^{\text{LO}} (\tilde{t}_2 \to A_1 \tilde{t}_1) = 9.26 \mbox{
  GeV} \;.
\eeq
In the whole investigated $m_{\tilde{t}_1}$ range, the EW and QCD
corrections are significant, 
decreasing together the LO width by $\sim12-15$\%. Both corrections are of similar
size, where, depending on the parameter point, once the QCD, once the
EW corrections are more important. At our starting parameter point
both corrections are almost equal resulting in a decrease of
\beq
\Delta \Gamma^{\text{QCD+EW}} = -11.7\% \;,
\eeq
{\it cf.}~Fig.~\ref{fig:stopdecays} (lower). 
Our results demonstrate that also in the stop decays both the QCD and
EW corrections have to be considered for a meaningful prediction of
the stop decay width. 

\section{Conclusions \label{sec:concl}}
The search for New Physics is one of the main tasks at the LHC. In the
absence of any direct sign of new resonances so far, the precise
investigation of the Higgs sector becomes more and more
important. Physics beyond the SM might reveal itself in modified Higgs
decay rates compared to the SM expectations or in the discovery of
additional Higgs bosons, unambiguous sign of a non-SM Higgs
sector. In view of the complexity of the experimental analyses and the
plethora of still possible New Physics extensions, it is evident that the
success of this research program depends on the precise predictions of
parameters and observables from the theory side. In this paper we have
calculated the NLO SUSY-EW and -QCD corrections to the
decays of a pseudoscalar NMSSM Higgs boson into stop pairs and of the
heavier stop into the lighter stop and a pseudoscalar Higgs boson. Both
processes can become important in certain regions of the parameter
space and hence contribute to the discovery channels of either the
pseudoscalar Higgs boson and/or the stop quarks. The NLO corrections turn out to
be important and, depending on the scenario, the EW corrections can be
of same size as the QCD corrections and also come with opposite
sign. Therefore not only the inclusion of higher order corrections is
important but also the consideration of both the QCD and the EW
corrections is indispensable for making reliable predictions. 

\subsubsection*{Acknowledgments}
JB, MM and KW have been supported in part by the DFG SFB/TR9
``Computational Particle Physics''  and furthermore JB in part by the
Institutional Strategy of the University of T\"ubingen (DFG,
ZUK~63). The authors thank Michael Spira for helpful discussions. 

\setcounter{equation}{0}
\section*{Appendix}
\begin{appendix}
\section{The Loop Functions \label{app:scalardef}}
The $D$ dimensional one-loop integrals encountered in the calculation
are the scalar one-, two- and three-point functions $A_0$, $B_0$ and $C_0$ as
well as the coefficient of the two-point tensor integral of rank one,
$B_1$. They are defined as
\beq
A_0 (m) &=& 16 \pi^2 \mu^{4-D} \int \frac{d^D q}{i (2\pi)^D}
\frac{1}{(q^2-m^2)} \\
B_0 (p^2; m_1, m_2) &=& 16 \pi^2 \mu^{4-D} \int \frac{d^D q}{i (2\pi)^D}
\frac{1}{(q^2-m_1^2)[(q+p)^2-m_2^2]} \\
C_0 (p_1^2, p_2^2, p_{12}^2; m_1, m_2,m_3) &=& 16 \pi^2 \mu^{4-D}
\int \frac{d^D q}{i (2\pi)^D} \times \nonumber \\
&& 
\frac{1}{(q^2-m_1^2)[(q+p_1)^2-m_2^2][(q+p_{12})^2-m_3^2]}  \\
p_\mu B_1 (p^2; m_1,m_2) &=& 16 \pi^2 \mu^{4-D} \int \frac{d^D q}{i (2\pi)^D}
\frac{q_\mu}{(q^2-m_1^2)[(q+p)^2-m_2^2]}\;,
\eeq
where 
\beq
p_{12} \equiv p_1 + p_2 
\eeq
with $p,p_1,p_2$ denoting the external momenta, which are taken as
incoming, and $m,m_1,m_2,m_3$ the masses of the loop particles. 
\end{appendix}

\vspace*{0.5cm}

\end{document}